\documentclass{llncs}

\usepackage{fullpage}
\usepackage{wrapfig}

\usepackage{ifpdf}
\usepackage{cite}
\usepackage[hidelinks]{hyperref}

\ifpdf
 \usepackage[pdftex]{graphicx}
 \graphicspath{{images/}}
 \DeclareGraphicsExtensions{.pdf,.jpeg,.png}
\else
 \usepackage[dvips]{graphicx}
 \graphicspath{{images/}}
 \DeclareGraphicsExtensions{.eps}
\fi

\usepackage{amsmath}
\usepackage{amsfonts}
\usepackage{algpseudocode}
\usepackage{subfig}
\usepackage{fixltx2e}
\usepackage{xspace}
\usepackage[percent]{overpic}

\usepackage[shortcuts]{extdash}

\usepackage{color}
{\makeatletter \gdef\xxx{\@ifnextchar[\xxx@lab\xxx@nolab}
  \long\gdef\xxx@lab[#1]#2{{\bf [\marginpar{xxx}{\sc #1}: \color{red}{#2}]}}
  \long\gdef\xxx@nolab#1{{\bf [\marginpar{xxx}#1]}}
}

\newcommand{\geosect}{\textsc{GeoSect}\xspace}
\newcommand{\GeoSectLocal}{\textsc{GeoSect\=/Local}\xspace}
\newcommand{\algo}{\textsc{LRM}\xspace}
\newcommand{\lmoves}{\mathcal{L}\xspace}
\newcommand{\sectorization}{\mathcal{S}\xspace}
\newcommand{\traffic}{\mathcal{T}\xspace}
\newcommand{\cost}{\mathrm{cost}\xspace}
\newcommand{\asp}{\textsc{ASP}\xspace}

\algnewcommand{\algorithmicgoto}{\textbf{go to}}
\algnewcommand{\Goto}[1]{\algorithmicgoto~\ref{#1}}

\title{Local Redesigning of Airspace Sectors}
\author{Irina Kostitsyna\thanks{Computer Science Department, Stony Brook University, {\tt ikost@cs.stonybrook.edu}} \and Joseph Mitchell\thanks{Department of Applied Mathematics and Statistics, Stony Brook University, {\tt jsbm@ams.sunysb.edu}}
\institute{}}

\begin{document}

\maketitle

\begin{abstract}
In this paper we study the \emph{Airspace Sectorization Problem} (\asp) where the goal is to find an optimal partition (sectorization) of the airspace into a certain number of sectors, each managed by an air traffic controller. The objective of the \asp is to find a ``well-balanced'' sectorization that distributes the workload evenly among the controllers. We formulate the \asp as a partitioning problem of a set of moving points in a polygonal domain. In addition to the requirement of balancing the workload, we introduce restrictions on the geometry of the sectorization which come from the Air Traffic Management aspects.

We investigate several versions of the problem that arise from different definitions of the notion of the workload and various choices of geometric restrictions on the sectorization. We conclude that most of the formulations of the problem, except maybe in some trivial cases,  are NP-hard.

Finally, we propose a \emph{Local Redesigning Method} (\algo), a heuristic algorithm that rebalances a given sectorization by adjusting the boundaries of the sectors. We evaluate \algo experimentally on synthetically generated scenarios as well as on the real historical traffic data. We demonstrate that the sectorizations produced by our method are superior in comparison with the current sectorizations of the US airspace.
\end{abstract}

\section{Introduction}

The current design of the \emph{National Airspace System} (NAS) was developed based on flight routes that were formed historically. Over the years, the demand and the geometry of the routes have changed dramatically, yet the NAS has undergone little change. As a consequence, the current sectorization is no longer able to accommodate the rapidly increasing demand.

The problem of designing a flexible and dynamic airspace architecture that is able to adapt to changing traffic flows is addressed by the \emph{Dynamic Airspace Configuration} (DAC) project as a part of the \emph{Next Generation Air Transportation System} (NextGen)~\cite{Lee08}. The NAS is partitioned into 22 Air Route Traffic Control Centers, each subdivided into sectors, which are overseen by air traffic controllers. The maximum workload that air traffic controllers can safely handle results in a limitation on the capacities of the sectors. If the changing traffic causes the demand on a sector to exceed its capacity, the sector will not be able to accommodate all the incoming flights. This will lead to some flights being rerouted around the congested area, and others to be delayed.

There are two basic approaches to handling changes in traffic. The first one is to design a new sectorization from scratch. Such methods concentrate on new traffic patterns, while discarding the old sectorization. Examples of this approach include: a cell-based Mixed Integer Programming (MIP) model~\cite{YD2004,Yousefi2005}; a sectorization method using binary space partitions~\cite{bms-gaoad-09,SYM2010}; a Voronoi diagram method~\cite{Xue2009}; a graph partitioning method~\cite{MCD07}, etc. While a clean-sheet sectorization design provides a wide range for finding an optimal solution, it is undesirable due to a human factor; it is important for controllers to be familiar with the geometry of sectors and traffic patterns.

The second approach is to perform a local rebalancing of the existing sectorization without introducing dramatic changes to the sectors. Klein \emph{et al.}~\cite{KRK2008} present an algorithm that shifts pre-specified thin subsectors between adjacent sectors to rebalance them. A local method for adjusting the boundaries of sectors that provides ``outs'' (or extra space) around weather constraints is proposed by Drew~\cite{Drew2010}. This method uses a force-based approach to adjust the boundaries in order to improve the capacities of the sectors that are most impacted by weather. The Voronoi method presented by Xue~\cite{Xue2009} includes a local rebalancing option as well as a clean-sheet design option; it adjusts the design of sectors by iteratively moving the Voronoi centers. The existing sectorization (the one created by applying the Voronoi method in clean-sheet mode) is used as the seed for a genetic algorithm that adapts the sectorization to the new demand. Local methods may change the topology of the design, including changes in the number of sectors. For instance, a pair of adjacent sectors may be combined into one, or a single sector may be split into two. Sector combination methods, based on computing predicted capacity gaps and then greedily combining pairs of sectors having the largest such gaps, have been proposed and examined in experiments of~\cite{BK2008}.

In this paper we discuss theoretical aspects of the \asp and present a new approach to the problem of redesigning sectorizations by local adjustments of the sector boundaries. We present a \emph{Local Redesigning Method} (\algo), a highly customizable multi-criteria optimization heuristic. We have developed \GeoSectLocal (a complement to \geosect sectorization tool introduced in~\cite{bms-gaoad-09, SYM2010}) that performs \algo adjustments to an input sectorization. The input sectorization can be any partition of the airspace region of interest, including current NAS sectorization, manually entered sectors, or the output of any other sectorization method, such as the top-down \geosect clean-sheet method.

\section{Problem Statement}

\begin{wrapfigure}{r}{0.4\columnwidth}
\centering
\includegraphics[width=0.38\columnwidth]{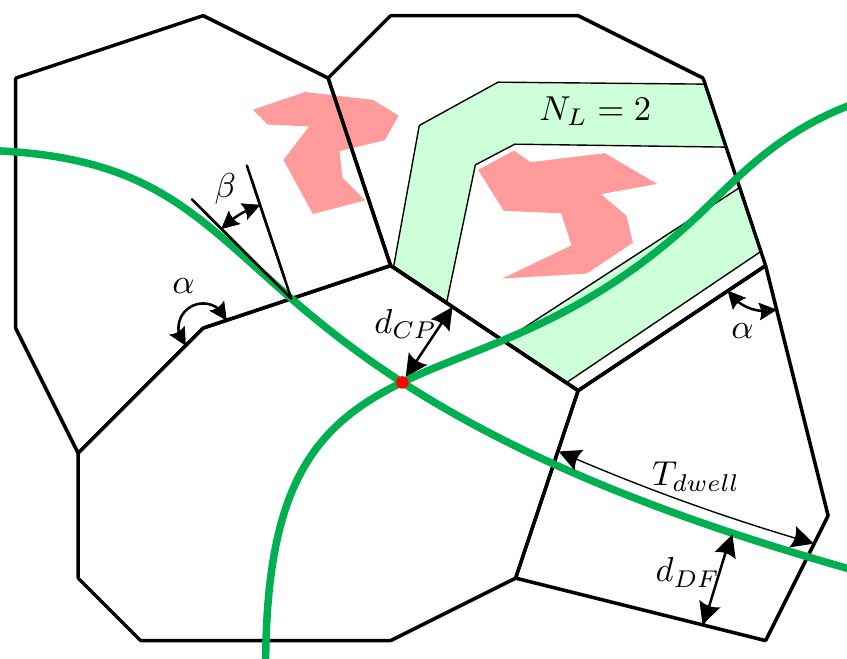}
\caption{\footnotesize Geometry of sectors, traffic flows}
\label{fig:requirements}
\end{wrapfigure}

The main objective of the \asp is to find a sectorization that distributes the workload evenly among the sectors. In this study we use three different metrics for the workload: $AC_{max}(\sigma)$---the maximum aircraft count in sector $\sigma$, $AC_{avg}(\sigma)$---the time-average aircraft count in $\sigma$, and $\delta(\sigma)$---the delay introduced by the overload of $\sigma$. We give precise definitions of $AC_{max}$, $AC_{avg}$ and $\delta$ in Section~\ref{subsec:obj_function}.

As discussed in~\cite{SYM2010}, human factor should also be taken into account when designing a sectorization. This leads to geometric restrictions that can be roughly divided into two groups: flow conformance requirements and sector geometry requirements. Flow conformance requirements reflect the compatibility of the sector boundaries with the traffic flows, weather obstacles, and locations of the airports or other singular points (refer to Figure~\ref{fig:requirements}). Airplanes should pass far enough from a sector boundary, and when they cross the boundaries between sectors the intersection angle should be nearly orthogonal. Furthermore, there is a restriction on the minimum dwell time for an airplane: after entering a sector the airplane should spend some time within it before exiting. Any \emph{critical point}, such as an airport or a conjunction/intersection of major flows, should also be well inside the sector to give air traffic controllers time to safely manage possible conflicts. If there is a weather obstruction in a sector, there should be enough of the throughput capacity in the directions of traffic flows to allow the sector to accommodate them. The second group of requirements regulates the geometry of sectors: we require the sectors to be convex or nearly convex and bound minimum and maximum angles.

Now, to formally state the \asp we can choose any of the discussed requirements to define the objective function $f(\cdot)$, and others to construct a set of constraints $\mathbb{C}$. For example, we can choose to optimize the average aircraft count subject to constraining sectors convexity and maximum aircraft count:
\begin{eqnarray*}
\text{optimize} & f&=\max_{\sigma}AC_{avg}(\sigma)\\
\text{subject to constraints} & \mathbb{C}&=\{\text{convex sectors}, AC_{max}(\sigma)\le AC^{*}_{max}\}\,.
\end{eqnarray*}

\begin{problem}[ASP]
Given a polygonal domain $\mathcal{D}$, a set of aircraft trajectories $\traffic$ in a time interval $[0,T]$, and, possibly, a set of dominant flows $\mathcal{DF}$, a set of critical points $\mathcal{CP}$, and a set of weather obstacles $\mathcal{W}$, find a partition $\sectorization$ of $\mathcal{D}$ into $k$ sectors to optimize $f(\sectorization,\traffic,\mathcal{CP},\mathcal{DF},\mathcal{W})$ subject to constraints $\mathbb{C}\,$.
\end{problem}

We also consider a dual problem whose goal is to minimize the number of sectors in a sectorization subject to a set of constraints.

\begin{problem}[dual ASP]
Given a polygonal domain $\mathcal{D}$, a set of aircraft trajectories $\traffic$ in a time interval $[0,T]$, and, possibly, a set of dominant flows $\mathcal{DF}$, set of critical points $\mathcal{CP}$, and a set of weather obstacles $\mathcal{W}$, find a partition $\sectorization$ of $\mathcal{D}$ into minimum number of sectors subject to constraints $\mathbb{C}$.
\end{problem}

\subsection{Complexity of \asp}

Few special cases of \asp can be solved polynomially. Consider \asp whose objective is to minimize the $AC_{avg}$. In the case when the airspace region is a convex polygon, this problem can be solved, for example, by line-sweeping technique. Balancing $AC_{avg}$ is equivalent to balancing the total length of aircraft trajectories in each sector. Sweep the line perpendicularly, slicing the polygon each time when a sector becomes ``full''. If the polygon is non-convex, this can result in sectors having multiple disconnected components, but this issue can be easily worked around by connecting the pieces with thin corridors along the polygon boundary.

Basu \emph{et al.}~\cite{bms-gaoad-09} prove that \asp whose objective is to minimize the maximum aircraft count with axis aligned rectangular sectors is NP-hard. Farrahi \emph{et al.}~\cite{Farrahi} strengthen this result by showing it to be NP-hard to optimize the maximum aircraft count version of the \asp with no geometric restrictions on sectors by reduction from the Planar-Partition-into-Triangles, or PLANAR-P3. They also show NP-hardness of the \asp version with the objective to minimize the total number of tracks in each sector, as well as the version with the objective to minimize the maximum number of aircrafts in a sector during any $N$-minute time interval (with a constant $N$).

\begin{wrapfigure}{r}{0.4\columnwidth}
\centering
\includegraphics[width=0.38\columnwidth]{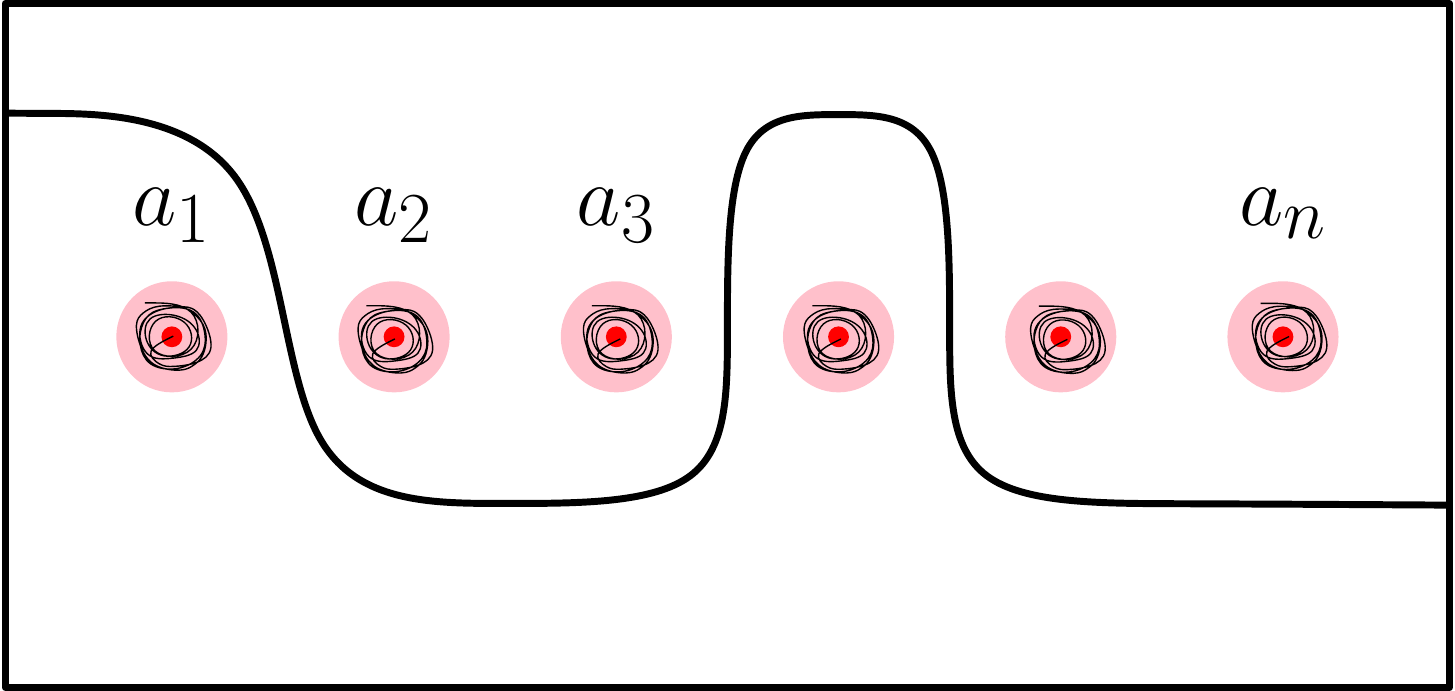}
\caption{\footnotesize An instance of the \asp with the constraint on the distance between critical points and sector boundaries constructed from an instance of the PARTITION problem. Red points represent airports with light-red disks showing the areas prohibited for sector boundaries to cross. Put $a_1,a_2\dots,a_n$ number of aircrafts to circle around the airports within the ``no\=/crossing'' areas.}
\label{fig:np_hard}
\end{wrapfigure}%
In this paper we show that a number of special cases of \asp are weakly NP-hard by a simple reduction from the PARTITION problem. The construction of the proof is basically the same for all the cases with slight differences that account for the types of the constraints. Here we show the proof that the \asp with the constraint on the distance between critical points and sector boundaries with the objective to minimize $AC_{avg}$ or $AC_{max}$ is weakly NP-hard. The PARTITION problem asks to determine if it is possible to divide a set of positive integers $a_1,a_2,\dots,a_n$ into two subsets such that the total sums of the numbers in each subset are equal. Given an instance of the PARTITION problem we construct an instance of the \asp with the required number of sectors $k=2$ (refer to Figure~\ref{fig:np_hard}). Any solution of the PARTITION problem corresponds to a subdivision of the airspace into two sectors that solves the \asp. Thus, we have the following theorem:
\begin{theorem}
\asp with the constraint on the distance between critical points and sector boundaries with the objective to minimize $AC_{avg}$ or $AC_{max}$ is weakly NP-hard.
\label{theorem:np_hard}
\end{theorem}

\begin{theorem}
\asp with the constraint on the minimum dwell time with the objective to minimize $AC_{avg}$ or $AC_{max}$ is weakly NP-hard.
\end{theorem}

\begin{theorem}
\asp with the constraint on the intersection angles of dominant flows and sector boundaries a with the objective to minimize $AC_{avg}$ or $AC_{max}$ is weakly NP-hard.
\end{theorem}

\begin{theorem}
\asp with the objective to minimize $AC_{max}$ is weakly NP-hard.
\end{theorem}

\section{Overview of the Local Redesigning Method}

As we have seen, solving the \asp optimally is difficult. We have developed a heuristic \algo that improves a given sectorization by locally adjusting sector boundaries. To estimate the ``quality'' of sectors of a sectorization before and after an adjustment, the \algo introduces an objective function $\cost(\sigma,\traffic,\mathcal{DF},\mathcal{CP},\mathcal{W})$ that depends on the constraints of the \asp. Here $\sigma$ is a sector, $\traffic$ is a set of aircraft trajectories, $\mathcal{DF}$ is a set of dominant flows, $\mathcal{CP}$ is a set of critical points, and $\mathcal{W}$ is a set of weather obstacles. We will denote the objective function as $\cost(\sigma)$ for short. The value of $\cost(\sigma)$ is zero if the constraints are satisfied, and it grows if the constraints get violated. We have extracted a set of elemental parameters, each of which can numerically measure some simple property of a sector in a given sectorization. These parameters include the maximum and time-average aircraft count, the estimated delay, the angles of sectors, the distances between critical points and sector boundaries, the intersection angles of traffic flows with the boundaries, etc. For each of the \asp constraints and a corresponding parameter we define a simple cost function that determines a penalty for the violation of the given constraint, the more the violation, the higher the penalty. We take a linear combination of these cost functions to be the objective function in the \algo:

\begin{equation}
\cost(\sigma)=\sum_{c\in\mathbb{C}}w_{c}\cost_{c}(\sigma)\,,
\end{equation}
where $c$ is a constraint, $\cost_{c}(\sigma)$ is a cost function associated with $c$, and $w_{c}$ is a user-specified constant.
Now the \algo can optimize an input sectorization with respect to this objective function. We will describe the parameters and the cost functions in more details in Section~\ref{subsec:obj_function}. 

At a high level, \algo redesignes a given sectorization with a sequence of local adjustments, each making small changes to the geometry of the sectors, \emph{e.g.}, by moving (or inserting/deleting) vertices or edges (see Section~\ref{subsec:local_moves}). Notice, that our method can be applied to 3D sectorizations as well, but here we focus on sectorizations that are constant with the altitude, so the problem can be viewed as planar partitioning.

Consider a sectorization, $\sectorization$, a planar partition of an airspace into sectors $\{\sigma_1,\sigma_2,\dots,\sigma_k\}$, where each $\sigma_i$ is a simple polygon. The boundary $\delta\sectorization$ of $\sectorization$ is the boundary of the region of interest. \algo adjusts the interior of $\sectorization$ leaving $\delta\sectorization$ unchanged. Let $\lmoves(\sectorization)$ be a set of feasible local adjustments to sectorization $\sectorization$ (the way these are selected is described in Section~\ref{subsec:local_moves}).

\begin{figure}[!th]
\begin{algorithmic}[1]
\Require Sectorization $\sectorization$ and traffic $\traffic$; and optionally set of dominant flows $\mathcal{DF}$, set of critical points $\mathcal{CP}$, and weather obstacles $\mathcal{W}$
\Ensure Sectorization $\sectorization^{*}$ which is in a local minimum with respect to the objective function $\cost(\sigma,\traffic,\mathcal{DF},\mathcal{CP},\mathcal{W})$
\Loop \label{marker}
\State define $\cost(\sigma)=\cost(\sigma,\traffic,\mathcal{DF},\mathcal{CP},\mathcal{W})$
\State $\mathbb{S} \leftarrow \{\sigma_1,\sigma_2\dots,\sigma_n\}$, sorted by $\cost(\sigma)$
\Loop
\If{$\mathbb{S} $ is empty}
\State \Return $\sectorization$
\EndIf
\State $\sigma_{max} \leftarrow \mathbb{S}.\mathrm{pop}()$
\State $c_0 \leftarrow \cost(\sigma_{max})$
\State $c_{\ell} \leftarrow \min_{\ell\in\lmoves(\sectorization)}\max_{\sigma\in\mathbb{S}}(\cost(\sigma_{\ell}))$
\If{$c_l>c_0$}
\State apply $\ell$ to $\sectorization$
\State \Goto{marker}
\EndIf
\EndLoop
\EndLoop
\end{algorithmic}
\caption{\footnotesize Local Redesigning Method optimizes sectorization $\sectorization$ with respect to the multi-criteria objective function. Here $\lmoves(\sectorization)$ is a set of feasible local adjustments to $\sectorization$ at every step of the algorithm, and $\sigma_{\ell}$ is a sector $\sigma$ after applying the local adjustment $\ell$.}
\label{alg:pseudocode}
\end{figure}

The objective function associates a penalty, $\cost(\sigma)$, to each sector $\sigma$. At each iteration of the main loop, we examine the sector, $\sigma_{max}$, with the highest cost. Consider $\lmoves(\sigma_{max})\subset\lmoves$: all the local adjustments that affect sector $\sigma_{max}$. \algo selects the best local adjustment $\ell_{max}\in \lmoves(\sigma_{max})$ that maximizes its ``benefit'', \emph{i.e.}, it minimizes the maximum cost among the sectors affected by $\ell(\sigma_{max})$, including $\sigma_{max}$ itself. If no $\ell_{max}$ can be selected from $\lmoves(\sigma_{max})$ that would decrease the maximum cost of sectors affected by $\lmoves(\sigma_{max})$, sector $\sigma_{max}$ is in the local minimum. In this case \algo moves to the sector with the next highest cost. Finally, after no further reduction in costs of the sectors is possible, the algorithm terminates, having found a locally optimal solution within the parameters of the search space. See the pseudocode in Figure~\ref{alg:pseudocode}.

We have  developed \GeoSectLocal, a tool implementing the \algo. As an input it takes a seed sectorization, scheduled traffic information, dominant flows and critical points, and a weather prediction. As an output, it produces a locally optimal sectorization with respect to the objective function. We use \geosect (presented in~\cite{SYK2010}) to extract dominant flows and critical points from the traffic schedule.

\subsection{Local Adjustments}
\label{subsec:local_moves}

\begin{figure}[!th]
\centerline{
  \subfloat[Cardinality and topology preserving]{
    \includegraphics[width=0.3\textwidth]{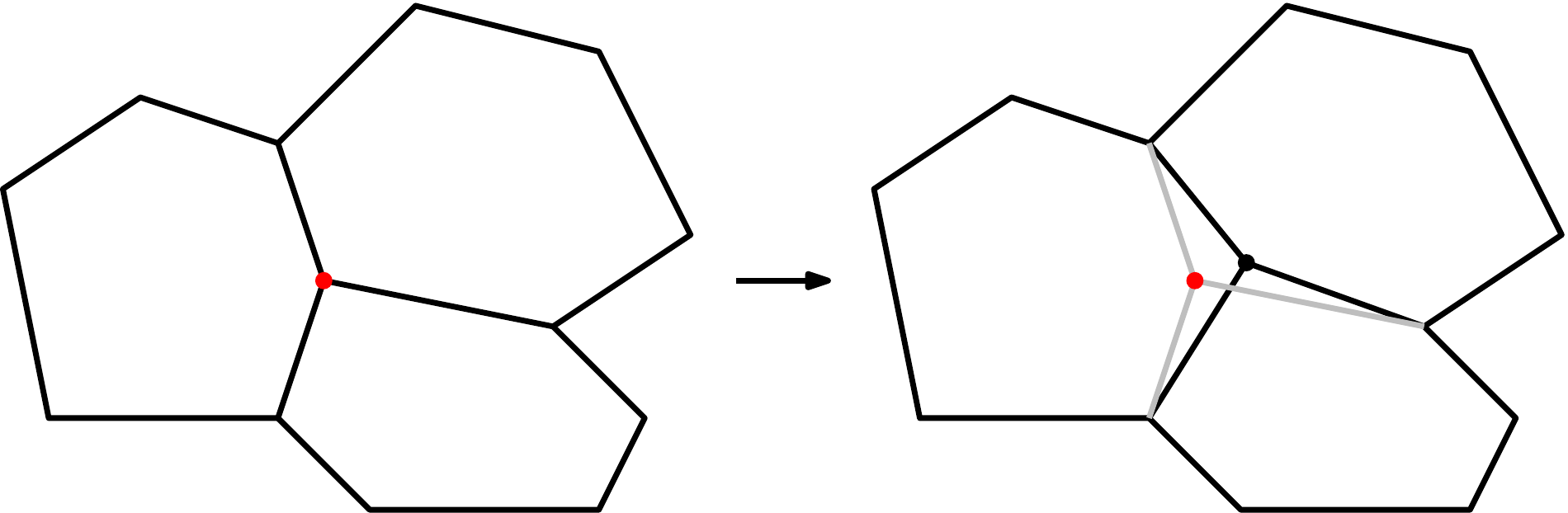}%
    \label{fig:local_move_1}}
  \qquad
  \subfloat[Cardinality preserving, topology modifying]{
    \includegraphics[width=0.3\textwidth]{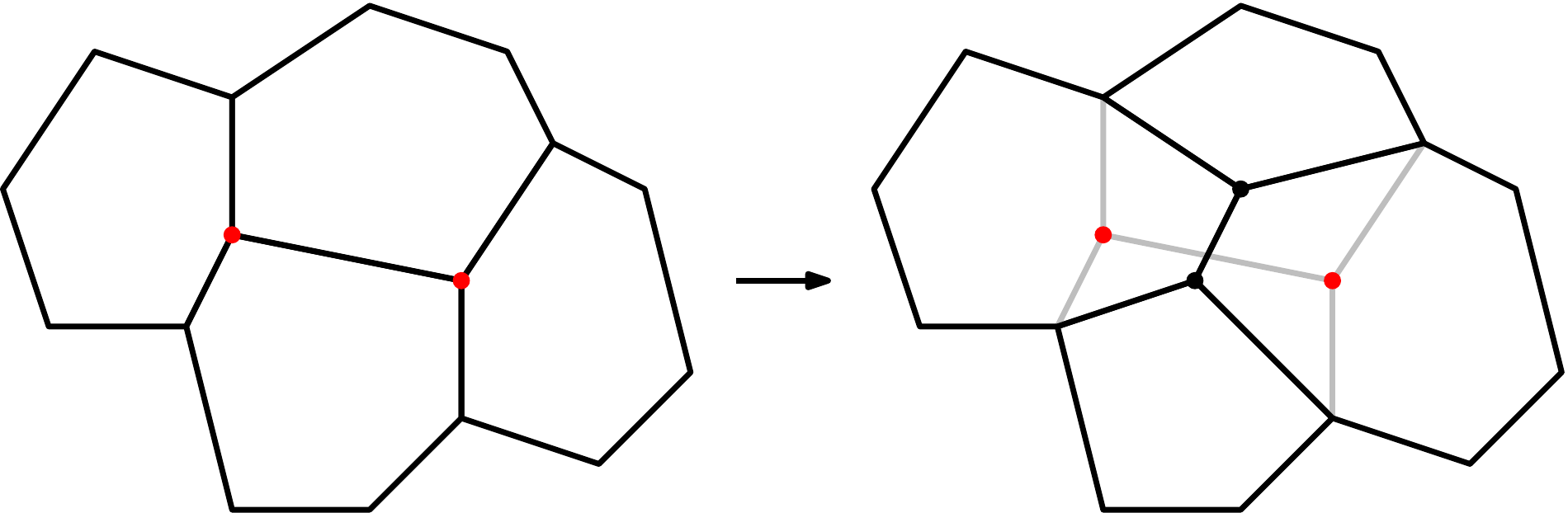}%
    \label{fig:local_move_2}}}
\centerline{
  \subfloat[Cardinality increasing, topology modifying]{
    \includegraphics[width=0.3\textwidth]{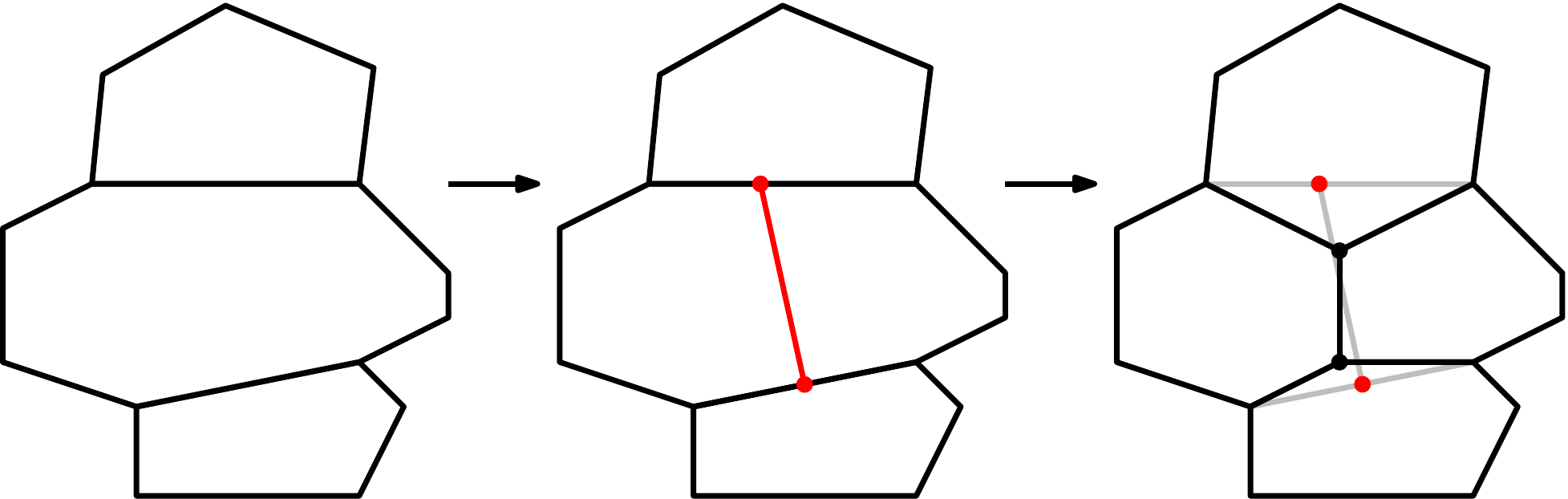}%
    \label{fig:local_move_3}}
  \qquad
  \subfloat[Cardinality decreasing, topology modifying]{
    \includegraphics[width=0.3\textwidth]{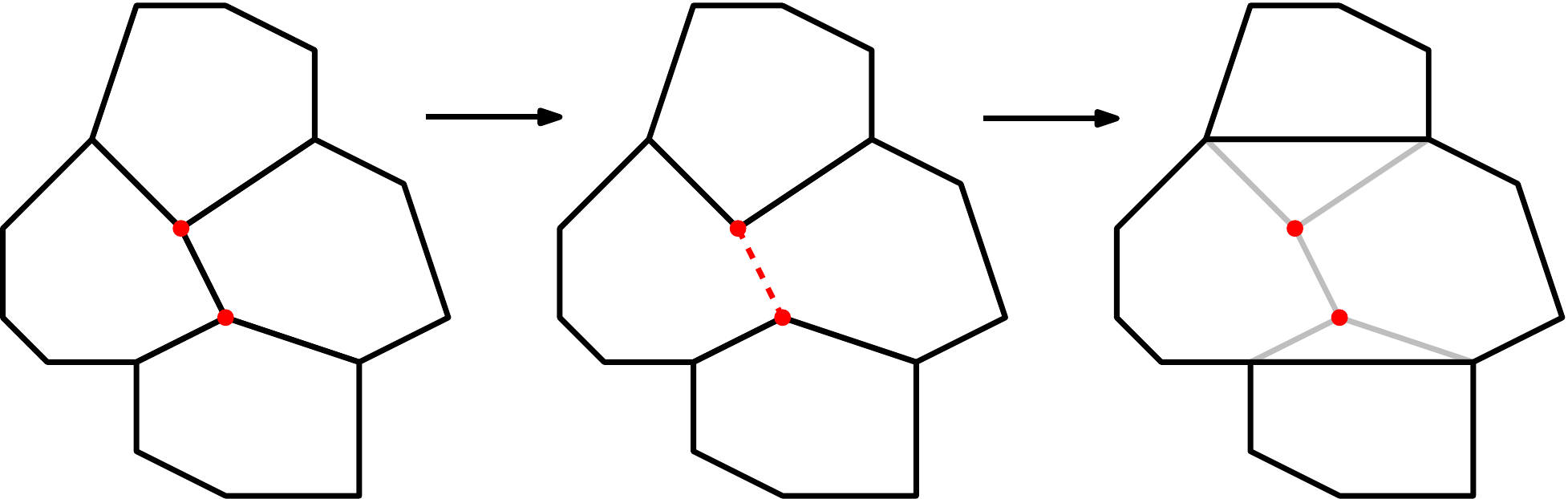}%
    \label{fig:local_move_4}}
    }
\caption{\footnotesize Examples of local adjustments. (a) and (b) are cardinality preserving; (c) and (d) are not.}
\label{fig:local_moves}
\end{figure}

We distinguish between different types of local adjustments to an airspace partition. We say that a set of local adjustments is cardinality preserving if the number of sectors remains the same after the adjustments. Cardinality preserving adjustments can be of two subtypes: topology preserving (the graph defining the partition remains constant, while the positions of the vertices may be changed), and topology modifying (allowing structural changes to the graph of sector boundaries). Figure~\ref{fig:local_move_1} is an example of a topology preserving adjustment, and Figure~\ref{fig:local_move_2} is an example of a topology modifying adjustment. Local adjustments that are not cardinality preserving are necessarily topology modifying (since the number of faces of the planar graph changes, implying a change in topology); such adjustments are of two subtypes: merge (two or more sectors are merged into one) and split (one or more sectors are split into a greater number of sectors). A simple example of a split operation is the partitioning of a sector into two or into three smaller sectors; after the split, the newly added vertices (each of degree 3) are readjusted, as in Figure~\ref{fig:local_move_3}. (Such partitions form the basis of \geosect's top-down recursive partitioning algorithm.) A simple example of a merge operation is the deletion of a shared boundary $uv$ in Figure~\ref{fig:local_move_4}, between two adjacent sectors; after deletion, vertices $u$ and $v$ (now each of degree 2, since they lie along sector boundaries) are adjusted so that the corresponding sector boundaries are reoptimized.

We implement two types of local adjustments in \GeoSectLocal: ``vertex move'' and ``edge flip''. These are cardinality preserving local adjustments (refer to Figure~\ref{fig:local_moves}). For ``vertex move'', we identify all possible locations for relocating a vertex $v$ using a grid centered around $v$. (Users can define the size and resolution of the search grid within the GUI.) The optimization evaluates the objective function at each candidate, and selects the candidate providing the lowest cost. For ``edge flip'' adjustment of edge $e$, we investigate edges that are perpendicular to $e$ and cross it in the middle. (Again, users can define the lengths and number of candidate edges.) While there are inefficiencies in this relatively brute-force approach to local optimization of vertex and edge position, using a search grid in this manner makes the testing of various objective functions straightforward and is consistent with the expectation that future airspace configurations will be based on an established grid of coordinates, the National Referencing System (NRS). We have tested some methods for reducing the number of discrete points searched for the optimal relocation, taking advantage of properties of the objective functions. However, these methods have had limited success. We have realized some computational efficiencies by evaluating components of the objective function only when necessary. Future work will address algorithm efficiency through selective search techniques, using bounding techniques to prune the search or using direct (non-grid) optimization techniques to optimize objective functions that are amenable to exact solutions.

\subsection{Details of the Objective Function}
\label{subsec:obj_function}

As we have mentioned before, we describe a set of elemental parameters that measure the ``quality'' of a sectorization. For each of the parameters $p$ we introduce one (or more) constraint $c$ and associate a penalty function $\cost_c(p)$. The penalty function $\cost_c(p)$ gives a measure of how far the parameter $p$ is outside of the permissible domain of values defined by $c$. For example, for parameter $\alpha$ (sector angle) one constraint can be to limit it from above by $\alpha_{max}$ to ensure that there will be no angles in sectors that are too large. Thus, $\cost_c(\alpha)$ measures how far $\alpha$ is above $\alpha_{max}$.

Figure~\ref{table:parameters} specifies all the constraints that are implemented in \GeoSectLocal. The definitions of the most of the parameters presented in the table are intuitive, however some of them require a discussion.

\paragraph{Maximum and time-average aircraft count.} Let $AC(\sigma, t)$ be the number of aircrafts in sector $\sigma$ at a time moment $t$. Then we can define the maximum aircraft count and the time-average aircraft count as
\begin{eqnarray}
AC_{max}(\sigma)&=&\max_{t}(AC(\sigma, t))\,,\\
AC_{avg}(\sigma)&=&\frac{\int AC(\sigma,t)dt}{\int dt}\,,
\end{eqnarray}

\paragraph{Estimated delay.} Let $K(\sigma)$ be the capacity of $\sigma$, \emph{i.e.}, the number of aircrafts that air traffic controllers can safely operate in sector $\sigma$. The capacity of the sector can be estimated using one of the following methods. One simple measure of $K$ commonly used is sector's MAP value, estimated by ($5/3$) times the average dwell time (in minutes). A more sophisticated estimate is that of~\cite{Welch2007}, which defines the maximum allowed aircraft count to be
\begin{equation}
K=\frac{-b+\sqrt{b^2-4ac}}{2a}\,,
\end{equation}
where $a=6.8/V$, $b=a+0.025+7/T$, $c=0.7$, $T$ is the average dwell time (in seconds), and $V$ is the sector volume (in cubic nmi). Now we can define an estimated the delay as
\begin{equation}
\delta(\sigma)=\int(AC(\sigma,t)-K(\sigma))dt\,.
\end{equation}%

\paragraph{Sector convexity.} We use a simple measure of non-convexity $cx(\sigma)$: the ratio of the area of sector $\sigma$ to the area of its convex hull, \emph{i.e.},
\begin{equation}
cx(\sigma)=\frac{\mathrm{area}(\sigma)}{\mathrm{area}(\mathrm{ConvexHull}(\sigma))}\,.
\end{equation}
If this ratio is $1$, the sector is convex. Smaller values of the ratio correspond to a greater degree of non-convexity.

\paragraph{Throughput.} We define a throughput value $N_L$ in a sector $\sigma$ along a dominant flows $df$ to be the number of air traffic lanes admissible by $\sigma$ along the flow $df$. The sectors with the throughput 2 or more provide additional alternate lanes that can be used for rerouting the traffic if needed. We compute $N_L$ by using the max-flow/min-cut analysis (refer to~\cite{Mitchell90,MPK06,MP07}).

\begin{figure}[!th]
\centering
\begin{tabular}{|c|c|l|c|c|c|}
\hline
\# & parameter & parameter description & constraint & default threshold & parameter limit\\
\hline
\hline
1 & $AC_{avg}$ & time-average airplane count & $AC_{avg}\le T_{1}$ & $T_{1}$ & $L_{1}=\infty$\\
\hline
2 & $AC_{max}$ & max airplane count  & $AC_{max}\le T_{2}$ & $T_{2}$ & $L_{2}=\infty$\\
\hline
3 & $\delta$ & estimated delay & $\delta\le T_{3}$ & $T_{3}=0$ & $L_{3}=\infty$\\
\hline
4 & $N_{L}$ & throughput & $N_{L}\ge T_{4}$ & $T_{4}=2$ & $L_{4}=0$\\
\hline
5 & $T_{dwell}$ & dwell time & $T_{dwell}\ge T_{5}$ & $T_{5}=300$ sec & $L_{5}=0$ sec\\
\hline
6 & $\beta$ & DF-boundary crossing angle & $\beta\le \beta_{max}$ & $T_{6}=\beta_{max}=45^{\circ}$ & $L_{6}=90^{\circ}$\\
\hline
7 & $d_{DF}$ & DF-boundary distance & $d_{DF}\ge T_{7}$ & $T_{7}=0.3''$ (long/lat) & $L_{7}=0''$\\
\hline
8 & $d_{CP}$ & CP-boundary distance & $d_{CP}\ge T_{8}$ & $T_{8}=0.4''$ (long/lat) & $L_{8}=0''$\\
\hline
9 & $\alpha$ & sector angle & $\alpha\ge\alpha_{min}$ & $T_{9}=\alpha_{min}=80^{\circ}$ & $L_{9}=0^{\circ}$\\
\hline
10 & $\alpha$ & sector angle & $\alpha\le\alpha_{max}$ & $T_{10}=\alpha_{max}=180^{\circ}$ & $L_{10}=360^{\circ}$\\
\hline
11 & $cx$ & sector convexity & $cx\ge cx_{min}$ & $T_{11}=cx_{min}=90\%$ & $L_{11}=0\%$\\
\hline
12 & $\left|e\right|$ & edge length & $\left|e\right|\ge T_{12}$ & $T_{12}=0.4''$ (long/lat) & $L_{12}=0''$\\
\hline
\end{tabular}
\caption{\footnotesize Description of constraints, corresponding parameters used in \algo, and default settings for \GeoSectLocal.}
\label{table:parameters}
\end{figure}%

Note that some of the objectives are defined either with respect to the set of all tracks (trajectories in the input data) or with respect to a set of \emph{dominant flows}, or route structures, given as input. The dominant flows represent the primary flows, or route structures, of traffic across the airspace of interest. Dominant flows are dynamic, and need to be updated over time with changing conditions in traffic, especially in cases of weather-impacted airspace. For related work on dominant flow extraction, see~\cite{SYK2010} and~\cite{SYM2010}.

\begin{figure}[!th]
\centering
\subfloat[The penalty function with $L>T$.]{
\begin{overpic}[width=0.2\textwidth,trim=0 -1em 0 0]{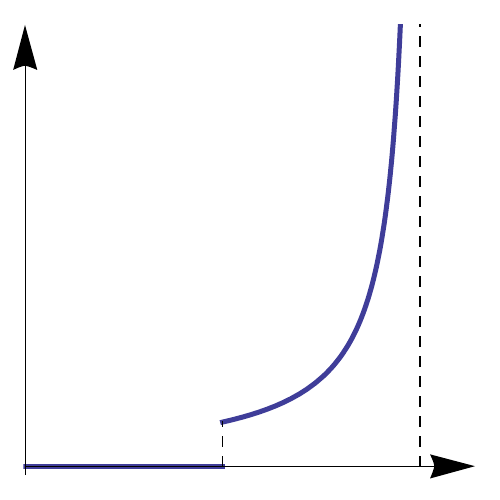}%
  \put(39,0){\small{$T$}}
  \put(77,0){\small{$L$}}
\end{overpic}
\label{fig:penalty_tl}}
\qquad
\subfloat[The penalty function with $T>L$. ]{
\begin{overpic}[width=0.2\textwidth,trim=0 -1em 0 0]{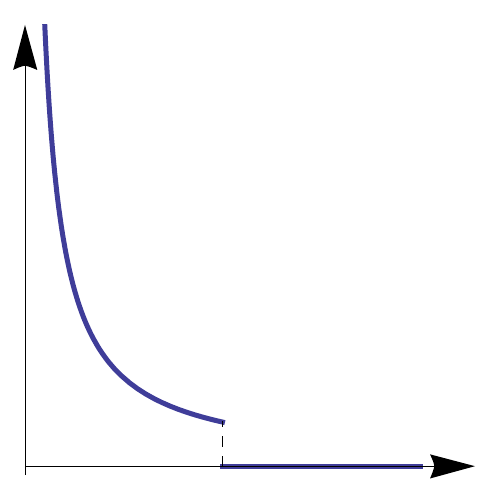}%
  \put(3,0){\small{$L$}}
  \put(39,0){\small{$T$}}
\end{overpic}
\label{fig:penalty_lt}}
\qquad
\subfloat[The penalty function with $L=\infty$.]{
\begin{overpic}[width=0.2\textwidth,trim=0 -1em 0 0]{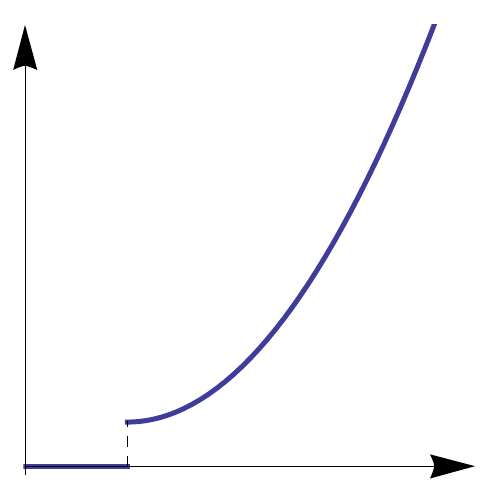}%
  \put(22,0){\small{$T$}}
\end{overpic}
\label{fig:penalty_inf}}
\caption{\footnotesize The penalty function.}
\label{fig:penalty_functions}
\end{figure}

The formulas of penalty functions $\cost_{c}(\sigma)$ depend on two constants $T_{c}$ and $L_{c}$, where $T_{c}$ is a user-defined threshold for parameter $p$, and $L_{c}$ is a physical limit for $p$ (refer to Figure~\ref{table:parameters}). For example, the upper limit of a sector angle is $360^{\circ}$ (sector angles cannot physically be greater than $360^{\circ}$), the lower limit of an edge length is $0$ (edges cannot have negative length), the upper limit for the average aircraft number in a sector is $\infty$ (technically, there is no limit on the number of airplanes in a sector). For the constraints on the parameters with bounded limit $L_{c}$ (such as constraints on sector angles, distances bounded from below, etc.) the penalty function is the following (refer to Figures~\ref{fig:penalty_tl} and~\ref{fig:penalty_lt}):
$$
\cost_{c}(p)=
\begin{cases}
 \displaystyle\frac{T_{c}-L_{c}}{p-L_{c}}\,, & \text{if }p\text{ is between }T_{c}\text{ and }L_{c}\,,\\
 0\,, & \text{otherwise.}
\end{cases}
$$
For the constraints on parameters with $L_{c}=\infty$ (see Figure~\ref{fig:penalty_inf}):
$$
\cost_{c}(p)=
\begin{cases}
 (p-T_{c})^{2}+1\,, & \text{if }p\ge T_{c}\,,\\
 0\,, & \text{otherwise.}
\end{cases}
$$

The choice of the objective function used in evaluating the quality of each of the candidate local adjustments is critical to the performance of the local optimization. One approach to multi-criteria optimization is to combine all the criteria into a single objective function, $\cost(\sigma)=\sum(w_c\cost_c(\sigma))$, according to user-specified weights $w_c$. We chose this approach in \GeoSectLocal as the most natural one.

Another approach would be to treat the individual components as a vector, $(c_1,c_2,\cdots,c_k)$, and optimize sets of components while constraining other components, \emph{e.g.}, to obtain Pareto-optimal solutions.

\section{Experimental Results}

\begin{figure}[!th]
\centering
\includegraphics[width=0.22\textwidth]{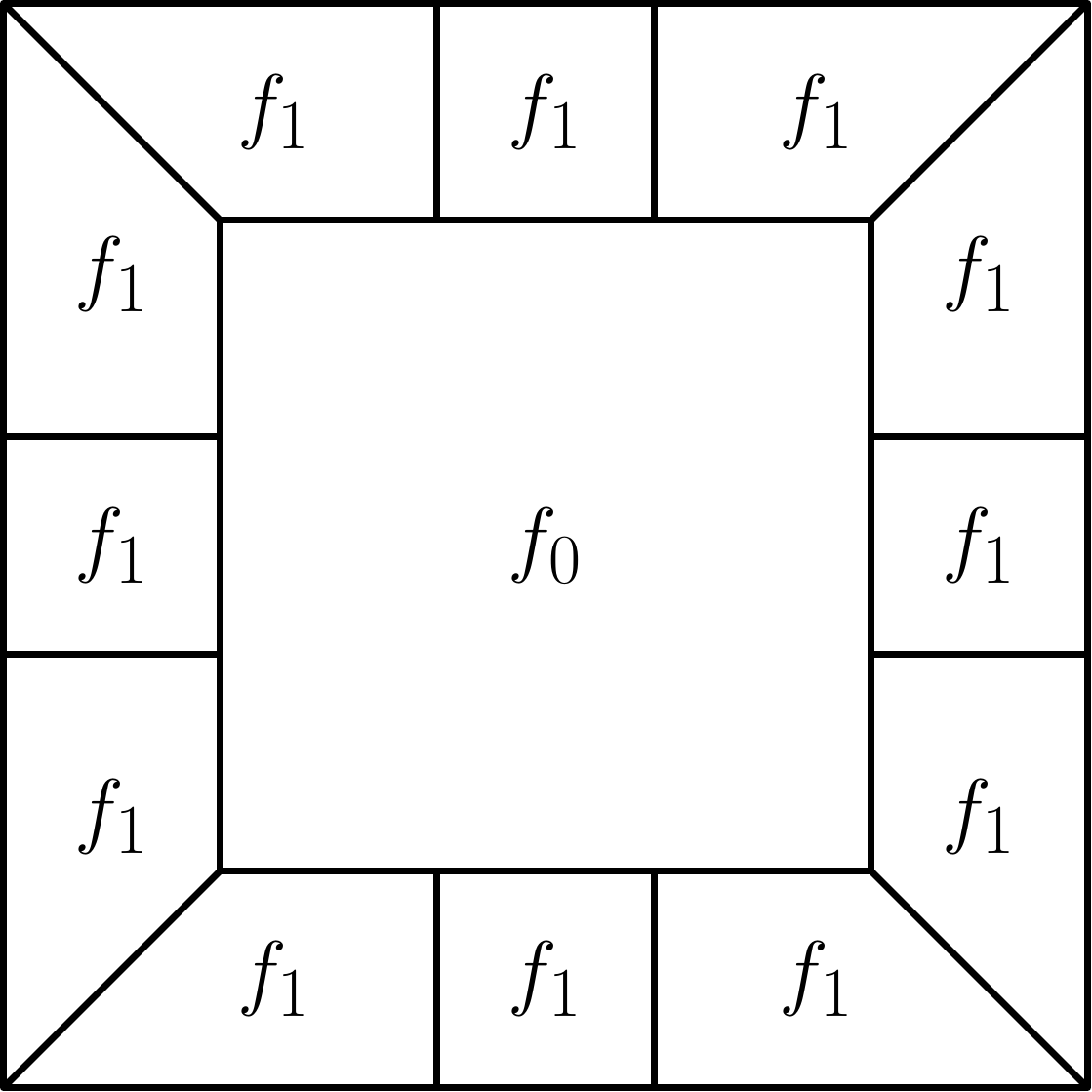}%
\qquad
\includegraphics[width=0.22\textwidth]{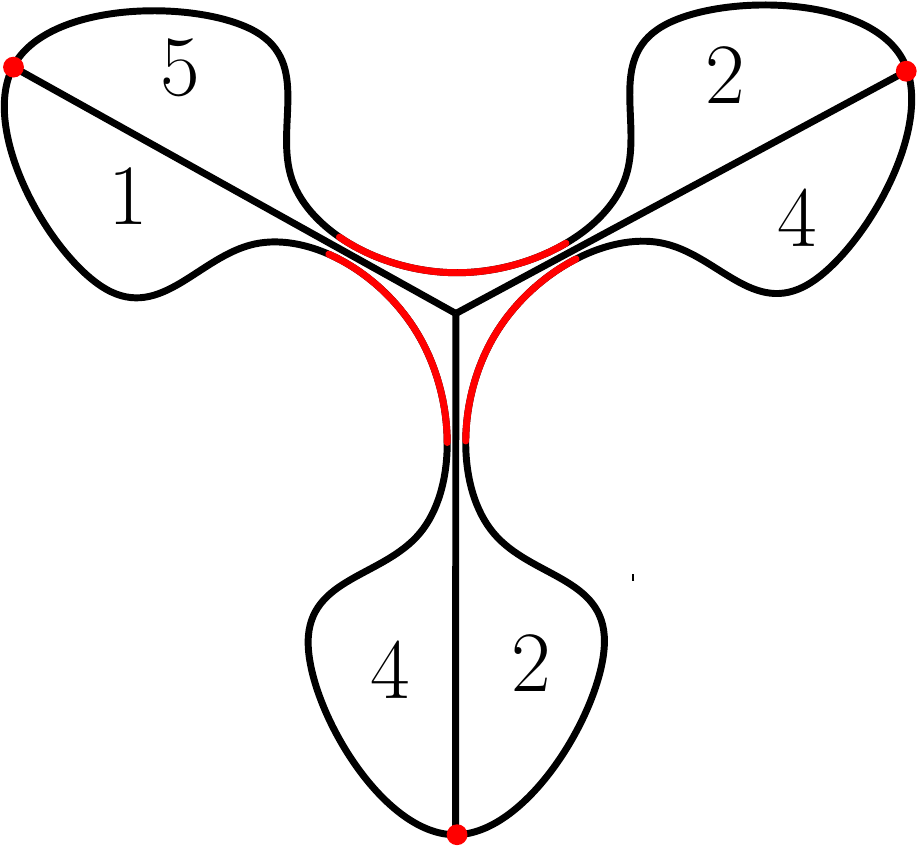}%
\caption{\footnotesize Bad examples when \GeoSectLocal can get stuck. Left: if $f_0<f_1$ are current sectors workloads, \GeoSectLocal cannot make a local adjustment while keeping sectors convex. Right: Let numbers in sectors denote the workload in the corresponding ``corner''. Optimum solution would be to connect the center with the red areas, but impossible to achieve by singular local moves.}
\label{fig:bad}
\end{figure}

We admit that in the worst case \algo can produce a sectorization that is arbitrarily bad in comparison with the optimal sectorization. For example, Figure~\ref{fig:bad} shows two toy situations when \algo can get ``stuck'' in a local minimum of the objective function. Left example shows a sectorization in a local minimum for a case when the weight of the penalty function of the convexity constraint is high enough in order not to allow appearance of non-convex sectors. Right example shows a case of a non-convex region that does not allow a single local adjustment to be made, and thus, is stuck in a local minimum.

Despite the obvious possibility of an arbitrarily bad solution, in practice \GeoSectLocal has proven to produce competitive sectorizations. In comparison with the sectorizations currently used by NAS \GeoSectLocal sectorizations improve the workload balance among sectors and reduce sector delays, in some cases by $50\%$ and more.

\subsection{Synthetic Experiments}

\begin{figure}[!th]
\centerline{
  \subfloat[Generated weather and traffic.]{
    \includegraphics[width=0.23\textwidth]{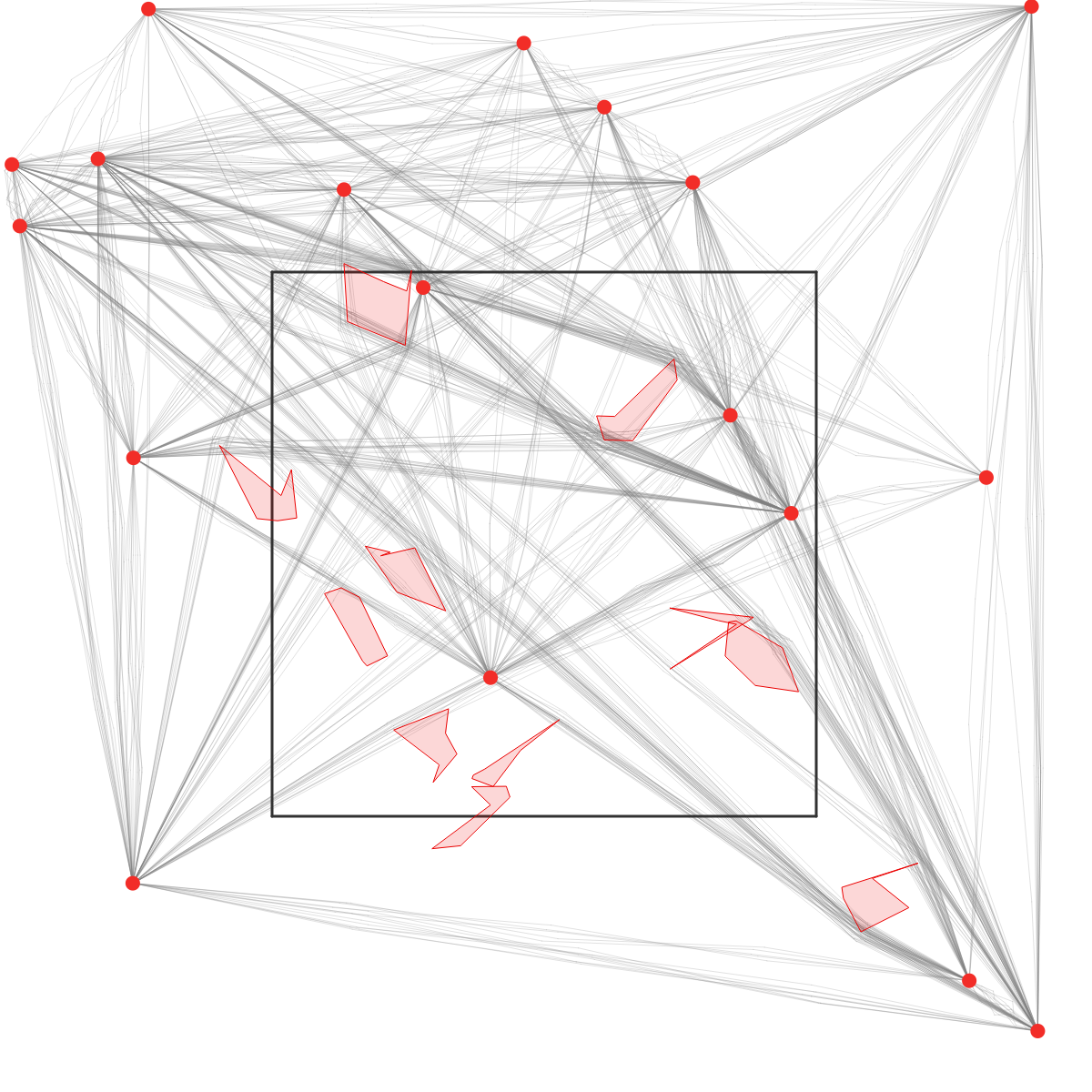}%
    \label{fig:synthetic_1}
    }
  \hfill
  \subfloat[Region of interest with extracted dominant flows.]{
    \includegraphics[width=0.23\textwidth]{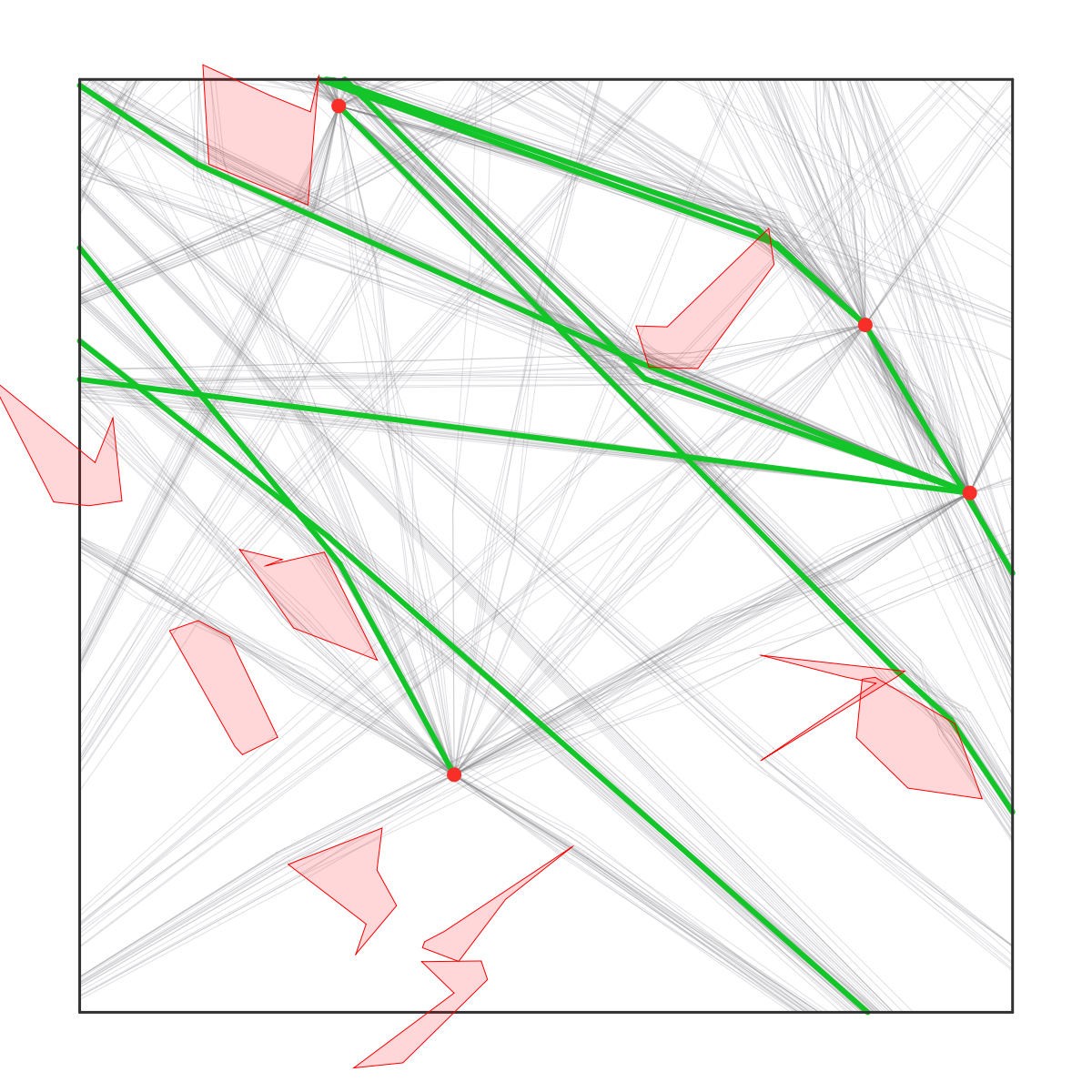}%
    \label{fig:synthetic_2}
    }
  \hfill
  \subfloat[Seed sectorization.]{
    \includegraphics[width=0.23\textwidth]{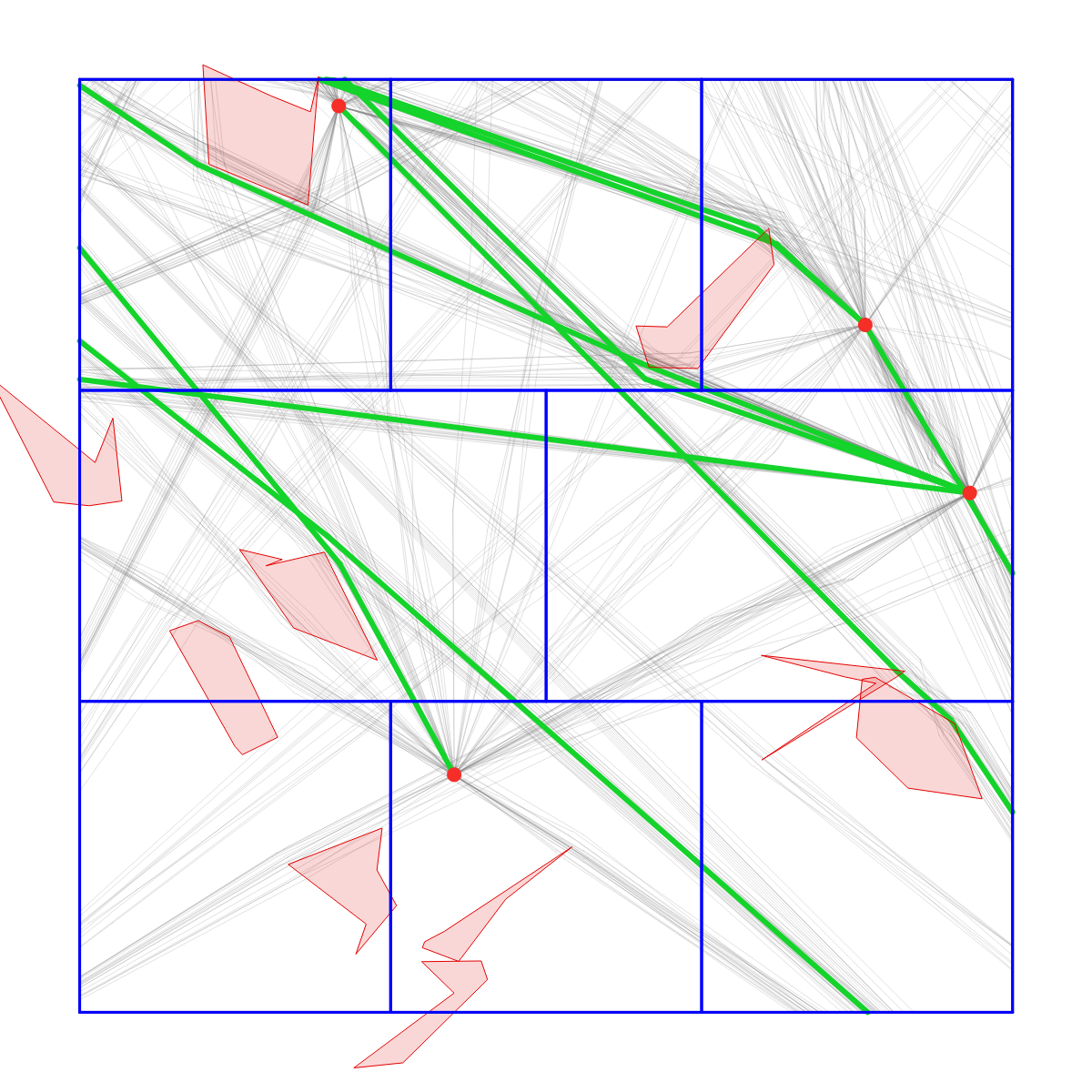}%
    \label{fig:synthetic_3}
    }
  \hfill
  \subfloat[\GeoSectLocal optimized sectorization.]{
    \includegraphics[width=0.23\textwidth]{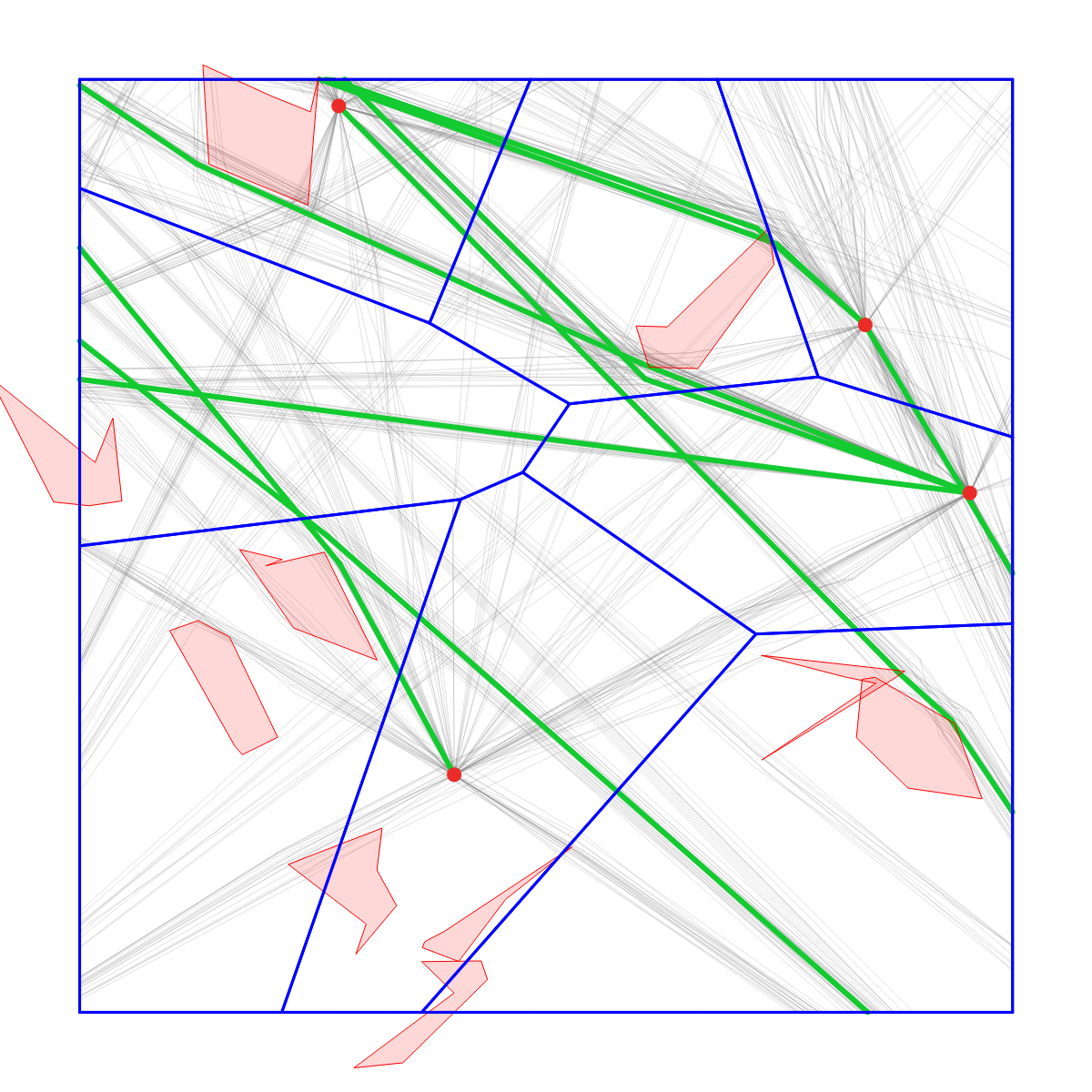}%
    \label{fig:synthetic_4}
    }}
\caption{\footnotesize Example of a synthetic experiment setting.}
\label{fig:synthetic}
\end{figure}

We have run a number of synthetic experiments to analyze the advantages and the drawbacks of the \algo. The test suit consisted of a number of randomly generated experiment settings consisting of a set of airports uniformly distributed in a $12''\times 12''$ square. Each airport had a corresponding normally distributed random weight which represented the size of the airport and affected the amount of traffic departing and landing at the airport. For each experiment setting we generated several random weather scenarios and sets of trajectories. The region that we chose to sectorize was a square of size $9''\times 9''$ in the center of the $12''\times 12''$ square. This way there was traffic passing through the region as well as fully contained inside it. Figure~\ref{fig:synthetic_1} shows one of the experiment settings with a weather scenario and a traffic generated for this scenario. Figures~\ref{fig:synthetic_2} and~\ref{fig:synthetic_3} show the region of interest with dominant flows extracted with the use of \geosect, and a sample seed sectorization. Finally, Figure~\ref{fig:synthetic_4} shows the output \GeoSectLocal sectorization.

\begin{figure}[!th]
\centerline{
\subfloat[Running time (sec) depending on the size of a search grid.]{
    \includegraphics[width=0.3\textwidth]{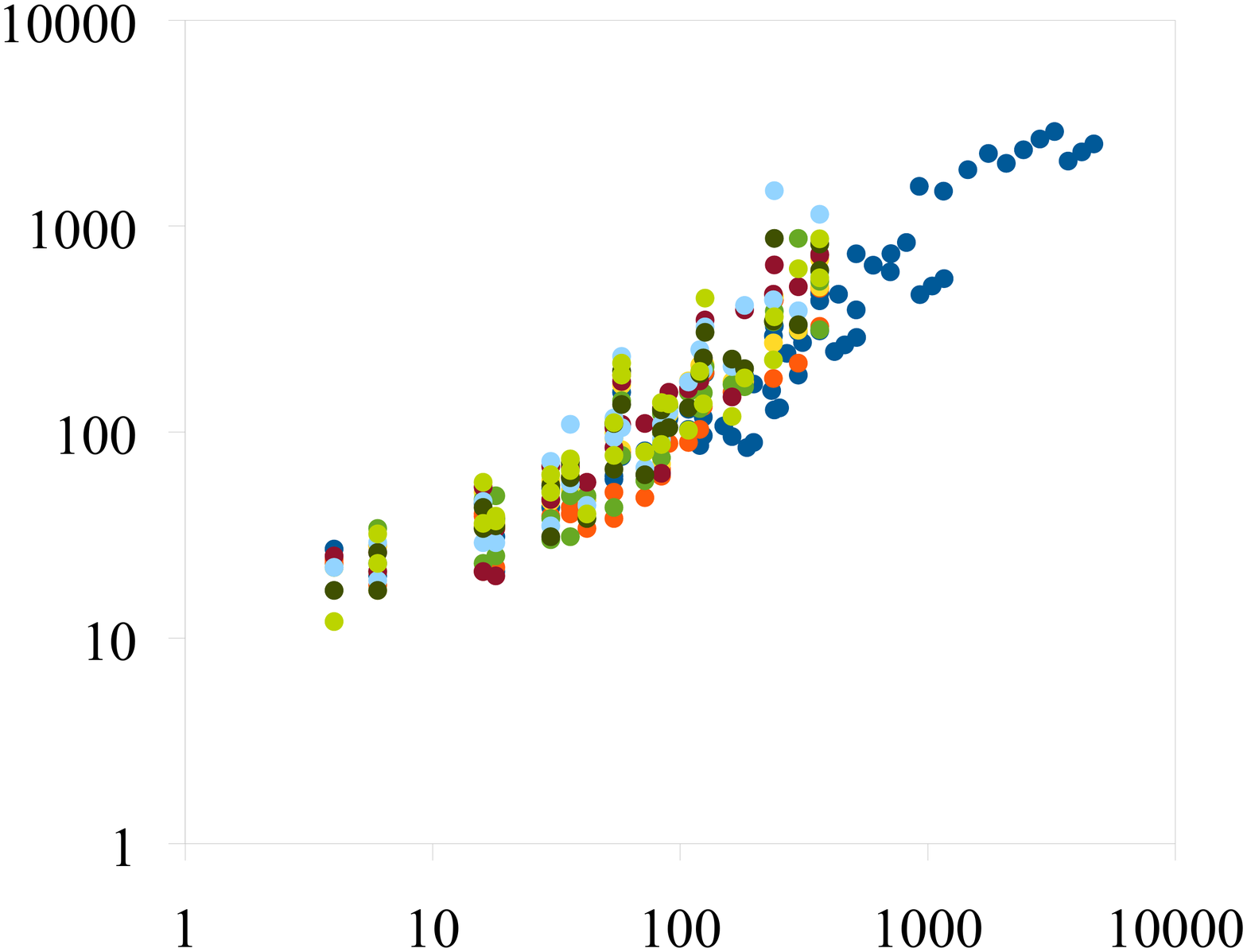}%
}
\qquad
\subfloat[Max value (over sectors) $AC_{avg}$ depending on the size of a search grid.]{
    \includegraphics[width=0.3\textwidth]{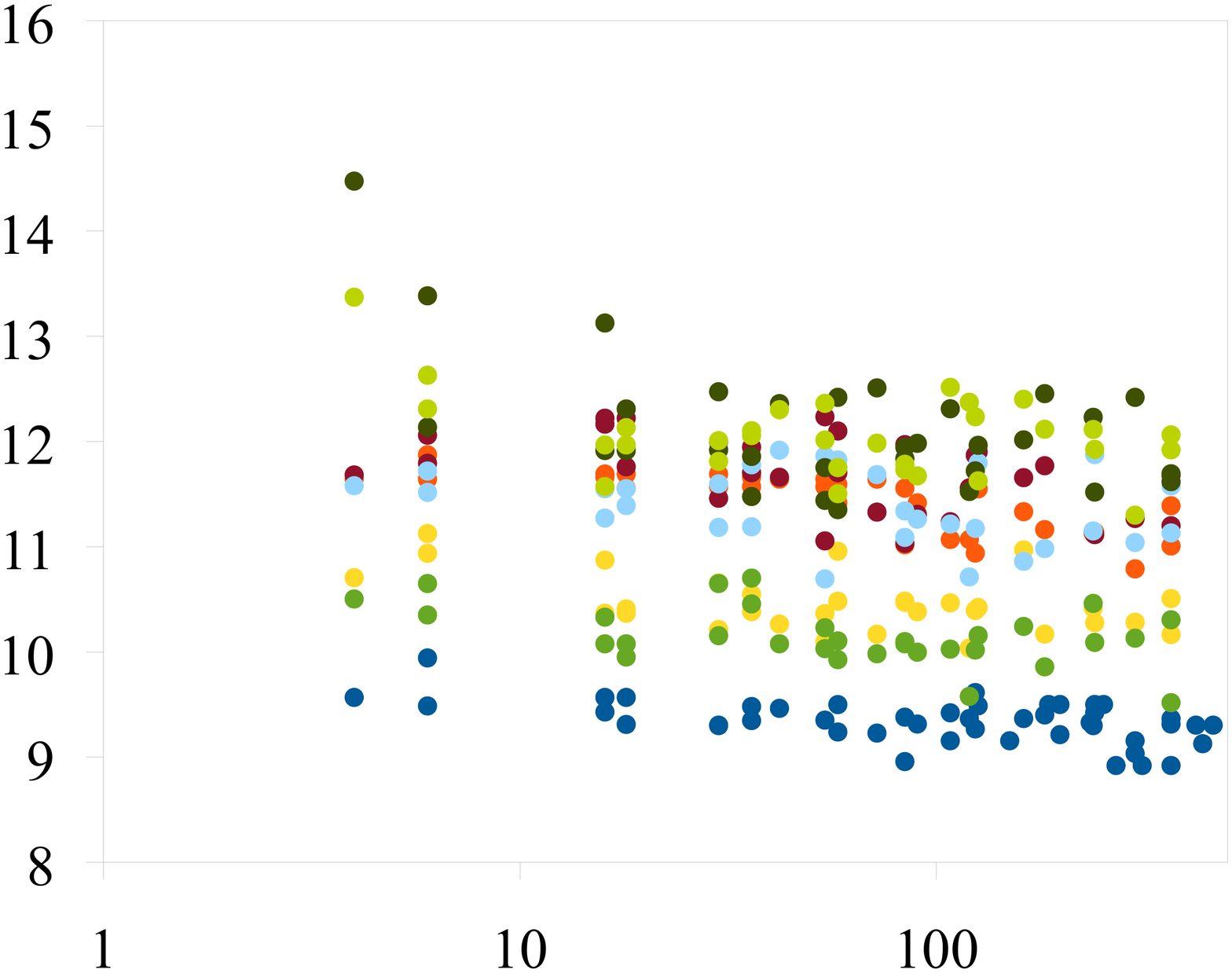}%
}
\qquad
\subfloat[StDev of the $AC_{avg}$ values depending on the size of a search grid.]{
    \includegraphics[width=0.3\textwidth]{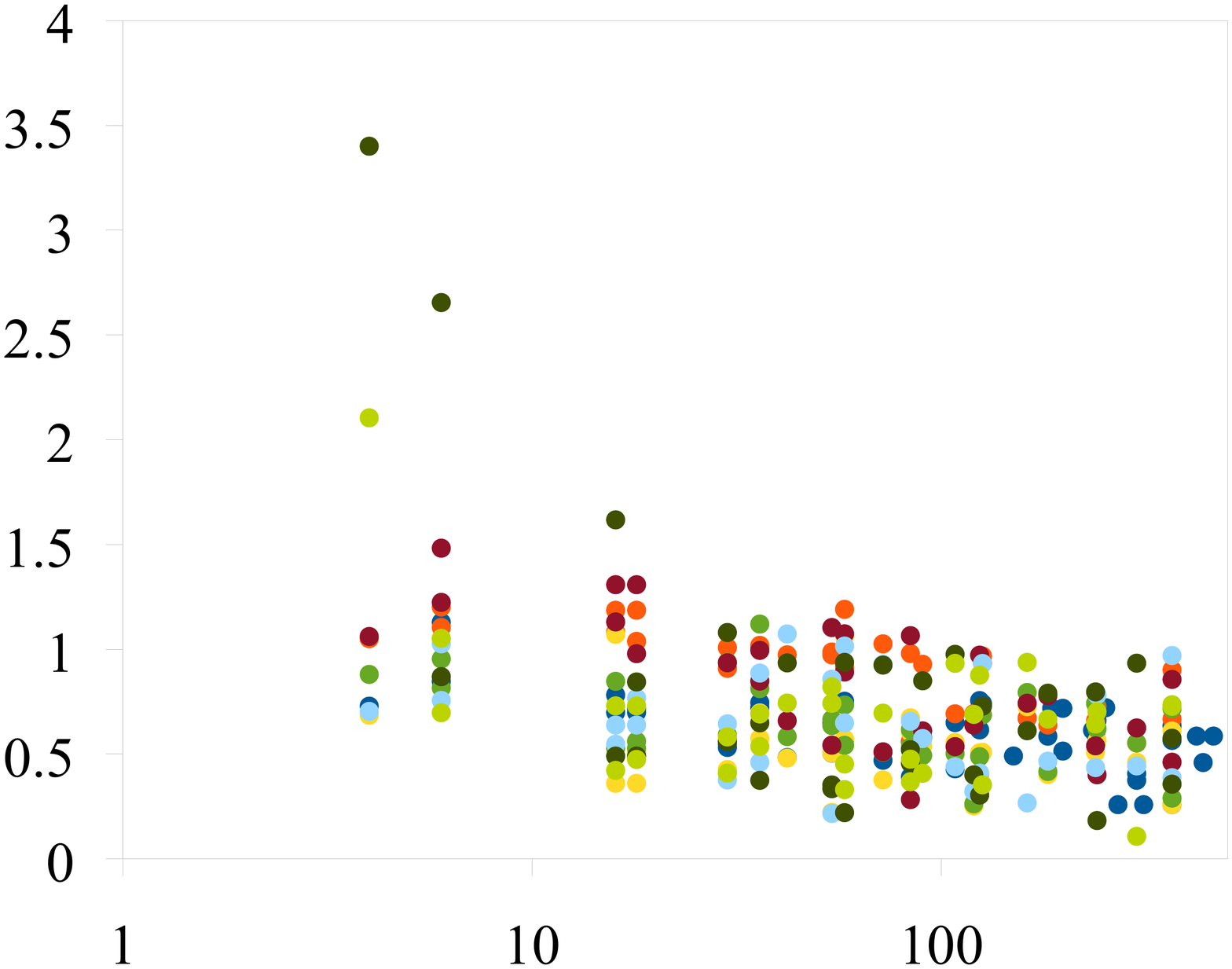}%
}}
\caption{\footnotesize Dependency of the running time and resulting $AC_{avg}$ on the grid size.}
\label{fig:first-round}
\end{figure}

The first round of experiments tested the dependency of the running time of the \algo and the variation of $AC_{avg}$ on the size of a search grid. The charts in Figure~\ref{fig:first-round} display the results for $8$ different experiment settings. We can see from these charts that the running time of the \algo nearly linearly depends on the size of the search grid, but there is not much of a gain in balancing the $AC_{avg}$ for grids with more than $30$ points. Based on this we chose the search grid to have $0.4''$ radius and $0.15''$ grid step (in total $36$ points) for the following experiments. 

\begin{figure}[!th]
\centering
\small
\begin{tabular}{|c|c|c|c|c|}
\hline
& \multicolumn{2}{|c|}{Worst $AC_{avg}$} & \multicolumn{2}{|c|}{Min convexity}\\
\hline
& AVG & StDev & AVG & StDev\\
\hline
\hline
No weather & 8.1 & 0.11 & 0.99 & 0.03\\
\hline
Weather 1 & 9.55 & 0.11 & 0.99 & 0.04\\
\hline
Weather 2 & 9.06 & 0.07 & 1 & 0\\
\hline
Weather 3 & 8.74 & 0.06 & 0.99 & 0.03\\
\hline
\end{tabular}
\caption{\footnotesize Dependency between the workload and the convexity.}
\label{fig:wl-convex}
\end{figure}

\begin{figure}[!tbh]
\centering
\subfloat[DF intersection angles (degrees) vs. $AC_{avg}$]{
    \includegraphics[width=0.3\textwidth]{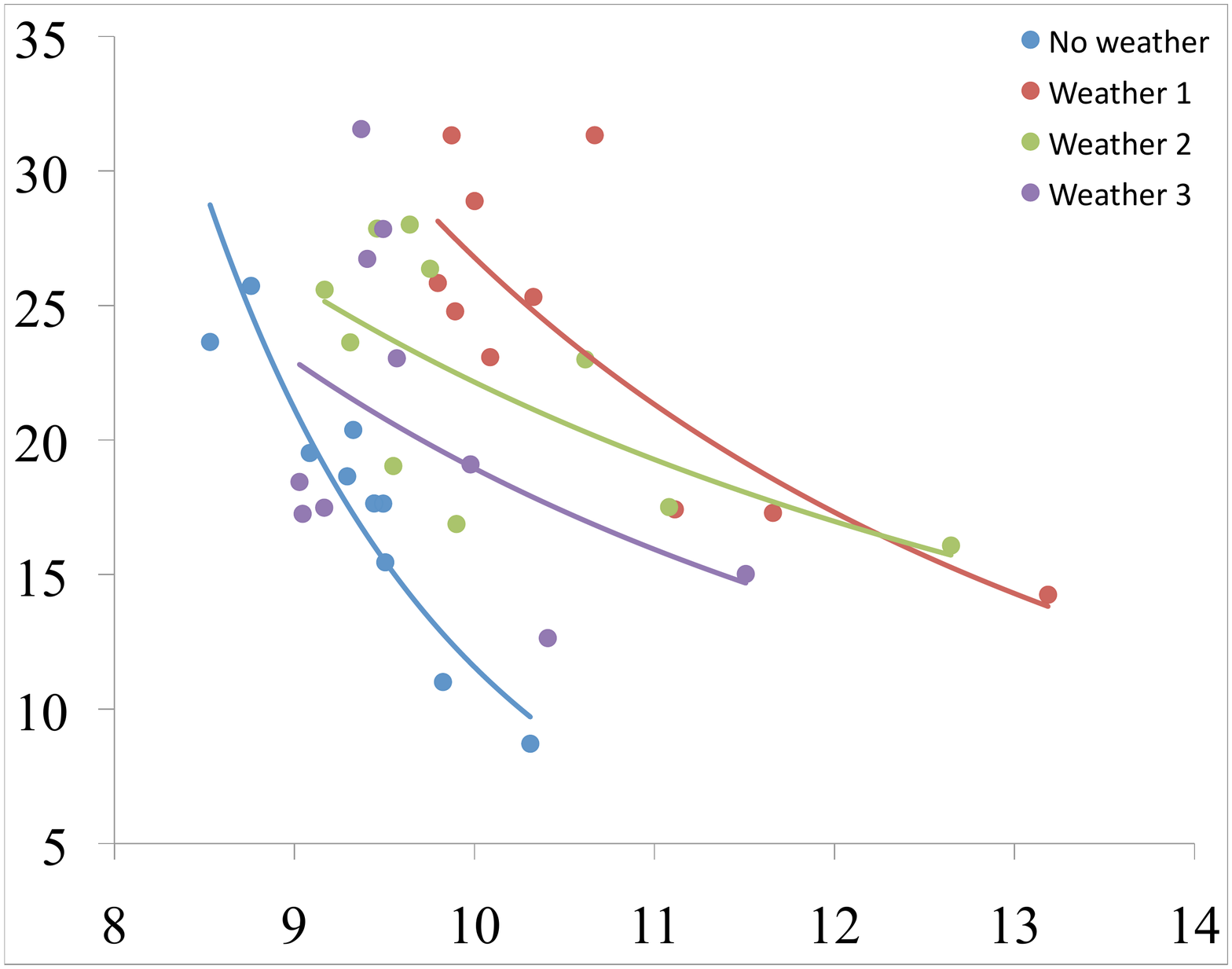}%
    \label{fig:wl-int-angle}
    }
    \hfill
\subfloat[DF dwell time (sec) vs. $AC_{avg}$]{
    \includegraphics[width=0.3\textwidth]{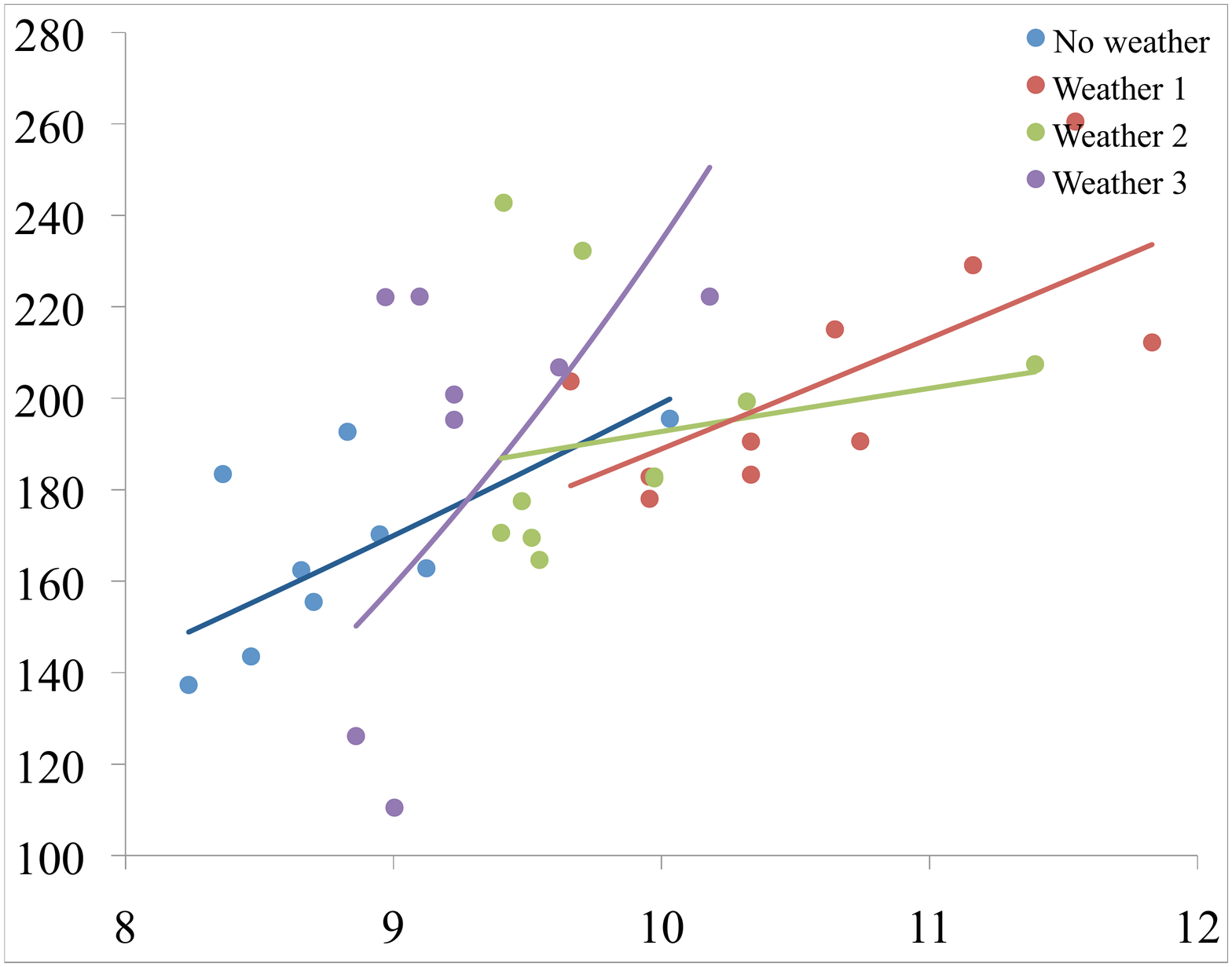}%
    \label{fig:wl-dwell}
    }
    \hfill
\subfloat[CP distance to the boundaries (degrees of long/lat) vs. $AC_{avg}$]{
    \includegraphics[width=0.3\textwidth]{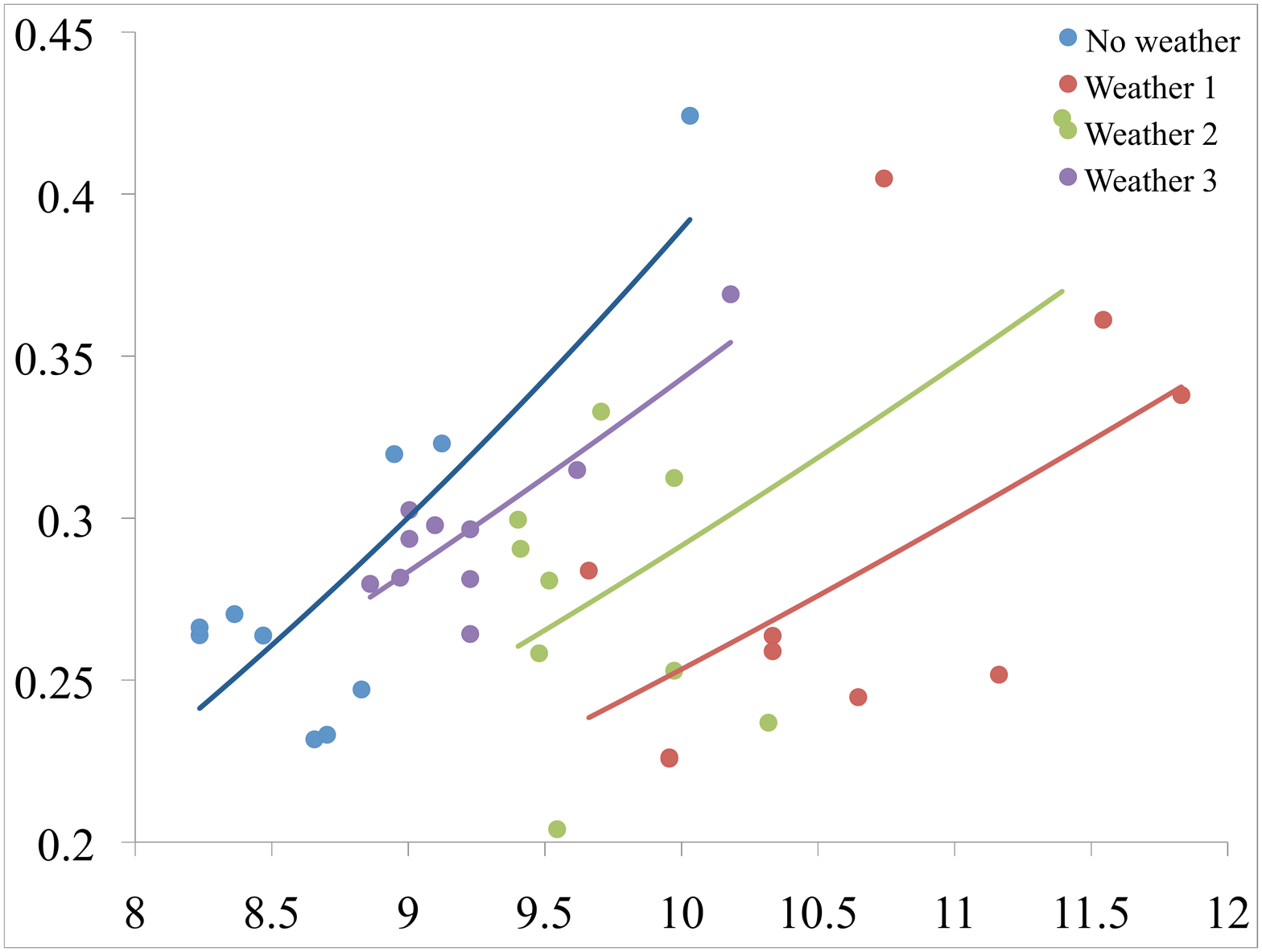}%
    \label{fig:wl-cp}
    }
\caption{\footnotesize Dependency between the workload and geometric parameters of sectorizations.}
\label{fig:param-dependency}
\end{figure}

We have run a number of tests to determine the correlation between the workload parameters and the geometric parameters. Figure~\ref{fig:wl-convex} shows a summary of an experiment where the $AC_{avg}$ was minimized, using various weight settings, under the convexity constraint. We observe that with no other constraints than the convexity constraint \GeoSectLocal produces sectorizations that are very close to optimal for any weight values.

We have also run similar experiments by selecting three other constraints to serve as a counterpart to the $AC_{avg}$ parameter: constraint on the maximum intersection angle of dominant flows and sector boundaries, constraint on the minimum dwell time, and constraint on the minimum distance between critical points and the boundaries. Figure~\ref{fig:param-dependency} shows the dependencies between the $AC_{avg}$ and these geometric parameters, that could be predicted: the smaller the average $AC_{avg}$ the higher intersection angles of the flows with the boundaries and the smaller minimum dwell time and the distance from the critical points to the boundaries.

\subsection{Historical Data Experiments}
\begin{figure}[!h]
  \centering
  \includegraphics[width=0.45\columnwidth]{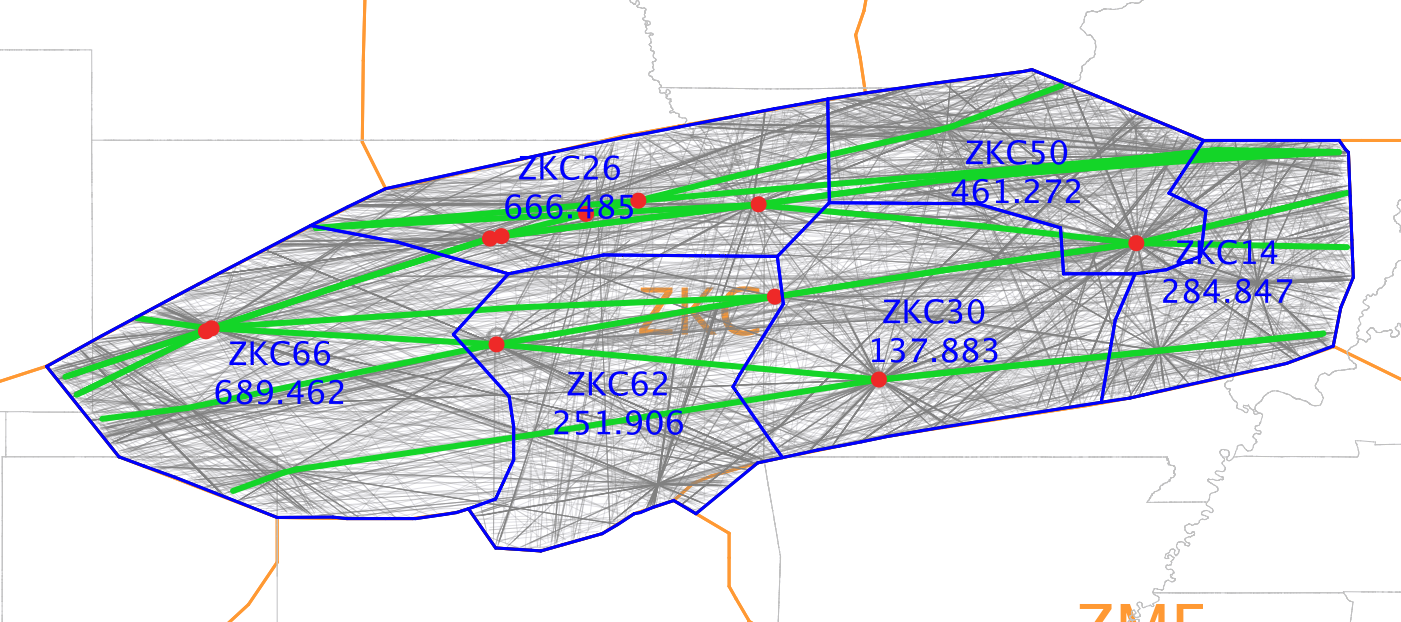}
  \qquad
  \includegraphics[width=0.45\columnwidth]{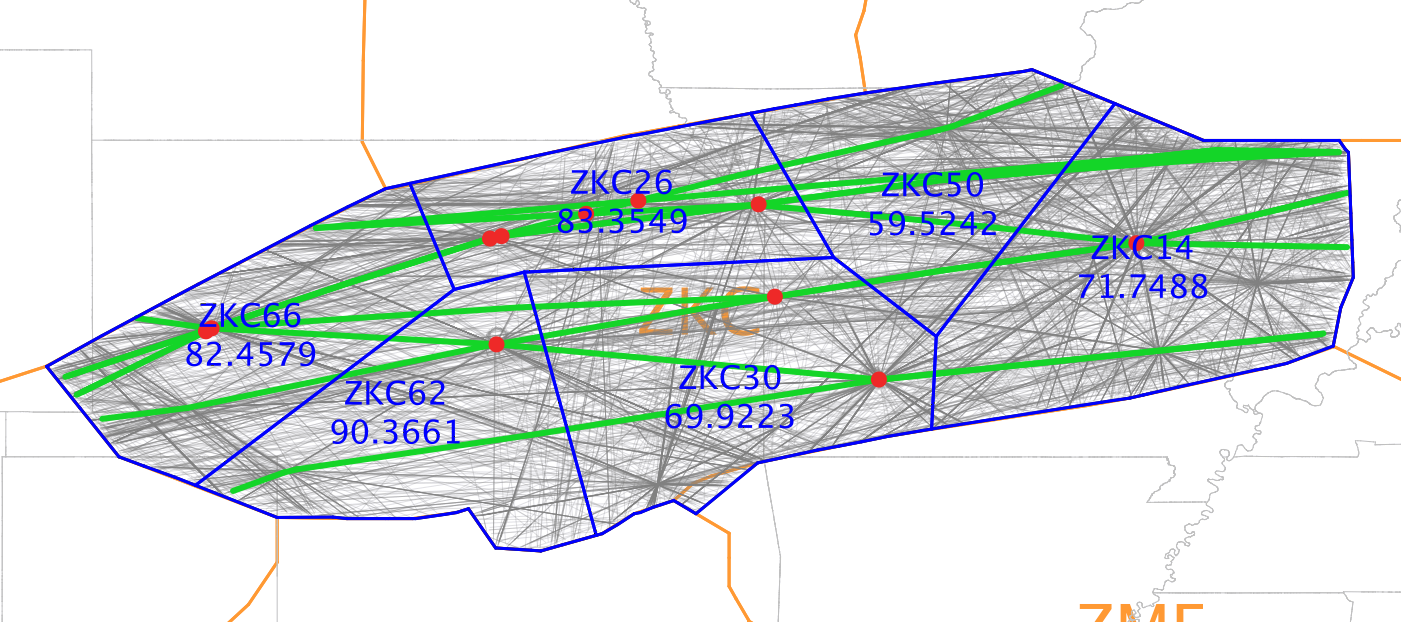}
  \caption{\footnotesize Baseline sectorization before and after rebalancing. Numbers in sectors show the total sector cost.}
  \label{fig:baseline}
\end{figure}

\begin{figure}[!h]
\centerline{
\small
\begin{tabular}{|l|c|c|c|c|c|c|c|c|}
\hline
 & \multicolumn{2}{|c|}{Parameter Value} & \multicolumn{6}{|c|}{Cost} \\
\hline
& \multicolumn{2}{|c|}{Worst} & \multicolumn{2}{|c|}{Avg} & \multicolumn{2}{|c|}{Max} & \multicolumn{2}{|c|}{StDev} \\
\hline
& before & after & before & after & before & after & before & after \\
\hline
\hline
min sec $\angle$ & $38^{\circ}$ & $59^{\circ}$ & 4.3 & 2.9  & 13.6 & 5.8 & 3.6 & 1.6 \\
\hline
max sec $\angle$ & $278^{\circ}$ & $191^{\circ}$ & 4.2 & 1.6 & 11.7 & 1.6 & 3.2 & 0 \\
\hline
convexity $cx$ & $0.85$ & $0.95$ & 1.5 & 0 & 1.5 & 0 & 0.03 & 0  \\
\hline
dist to CPs & $0.08''$ & $0.4''$ & 35.5 & 7.3 & 138.7 & 9.6 & 51.0 & 1.7 \\
\hline
dist to DFs & $0.08''$ & $0.25''$ & 49.9 & 12.5 & 103.1 & 20.0 & 30.0 & 5.2 \\
\hline
max $\angle$ w/ DFs & $76^{\circ}$ & $65^{\circ}$ & 23.1 & 9.8 & 64.2 & 23.8 & 24.1 & 8.2 \\
\hline
min dwell & $42$ sec & $125.5$ sec & 51.9 & 40.5 & 167.1 & 40.5 & 77.2 & 0 \\
\hline
est. delay & $20.4$ min & $15.3$ min & 105.4 & 35.4 & 278.9 & 62.7 & 104.5 & 23.3 \\
\hline
total cost: & \multicolumn{2}{|c|}{} & 415.3 & 76.2 & 689.5 & 90.4 & 228.5 & 11.2 \\
\hline
\end{tabular}
}
\caption{\footnotesize Comparisons of the parameters and penalties of the baseline sectorization before and after rebalancing.}
\label{fig:costs}
\end{figure}
In the experiment presented in this section we concentrate on minimizing the average delay in the Kansas City (ZKC) center for 36 hours of historical traffic data. We set the priority of the optimization to be the estimated delay by putting higher weight on the corresponding penalty function and lower weights on other constraints. As an input sectorization to \GeoSectLocal we selected the current NAS sectors for the ZKC center. Fig.~\ref{fig:baseline} shows the baseline sectorization before and after rebalancing.

Figure~\ref{fig:costs} compares parameters' penalties for the baseline sectorization (before and after rebalancing). There are improvements in average and maximum values as well as in standard deviation of penalties for all the parameters. \GeoSectLocal improved the average delay over all sectors and reduced the maximum delay by 25\%.

\subsection{Simulation Experiment}
\label{subsection:aces}

In collaboration with NASA Ames and Metron Aviation, we conducted experiments utilizing NASA's ACES Flight simulator. The flight simulator takes scheduled flights and a sectorization as an input and generates flight trajectories.

The goal of the experiment was to evaluate the sectorizations produced be \algo in ``real-life'' settings of ZKC center. The experiment had 3 stages corresponding to 3 time periods in a day: early morning (light traffic), early afternoon (the heaviest traffic), and evening (normal traffic). There are 3 current baseline sectorizations corresponding to these time intervals consisting of 6, 24 and 19 sectors respectively. A set of sample trajectories (for flights schedule with twice the usual demand) was generated by the simulator for the first stage with the option no input sectorization. Based on this projected flight trajectories for the first time interval we were to produce a sectorization and feed it in the flight simulator for the generation of the flight trajectories for the second stage of the experiment. The output sectorization of the second stage was used in the same way to generate the trajectories for the third stage.

\begin{figure}[!th]
\centerline{
\subfloat[Average flight delay (minutes).]{
    \includegraphics[width=0.32\textwidth]{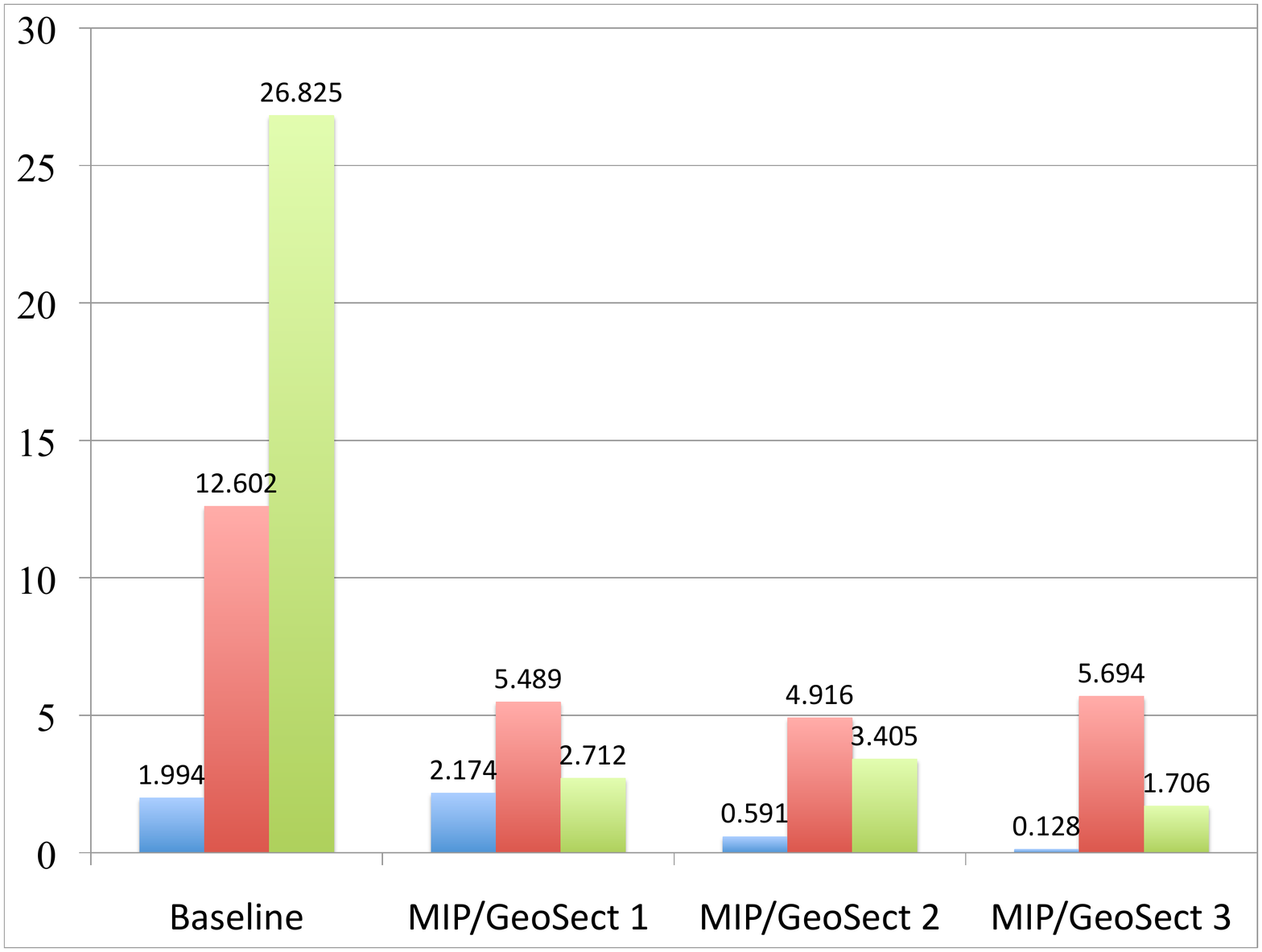}%
    \label{fig:aces_avg_wl}
}
\hfill
\subfloat[Maximum flight delay (minutes).]{
    \includegraphics[width=0.32\textwidth]{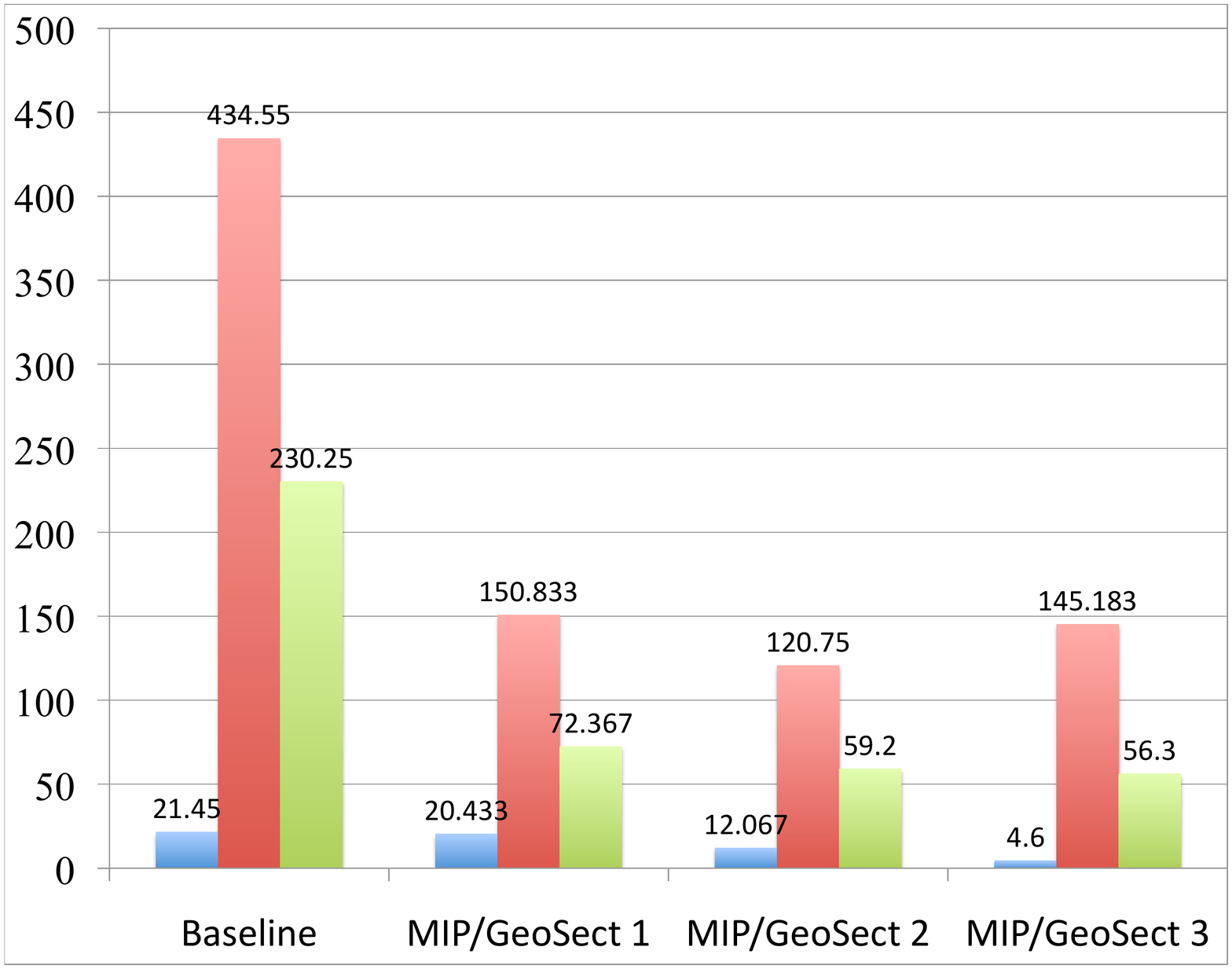}%
    \label{fig:max_wl_chart}
}
\hfill
\subfloat[Standard deviation of flight delay (minutes).]{
    \includegraphics[width=0.32\textwidth]{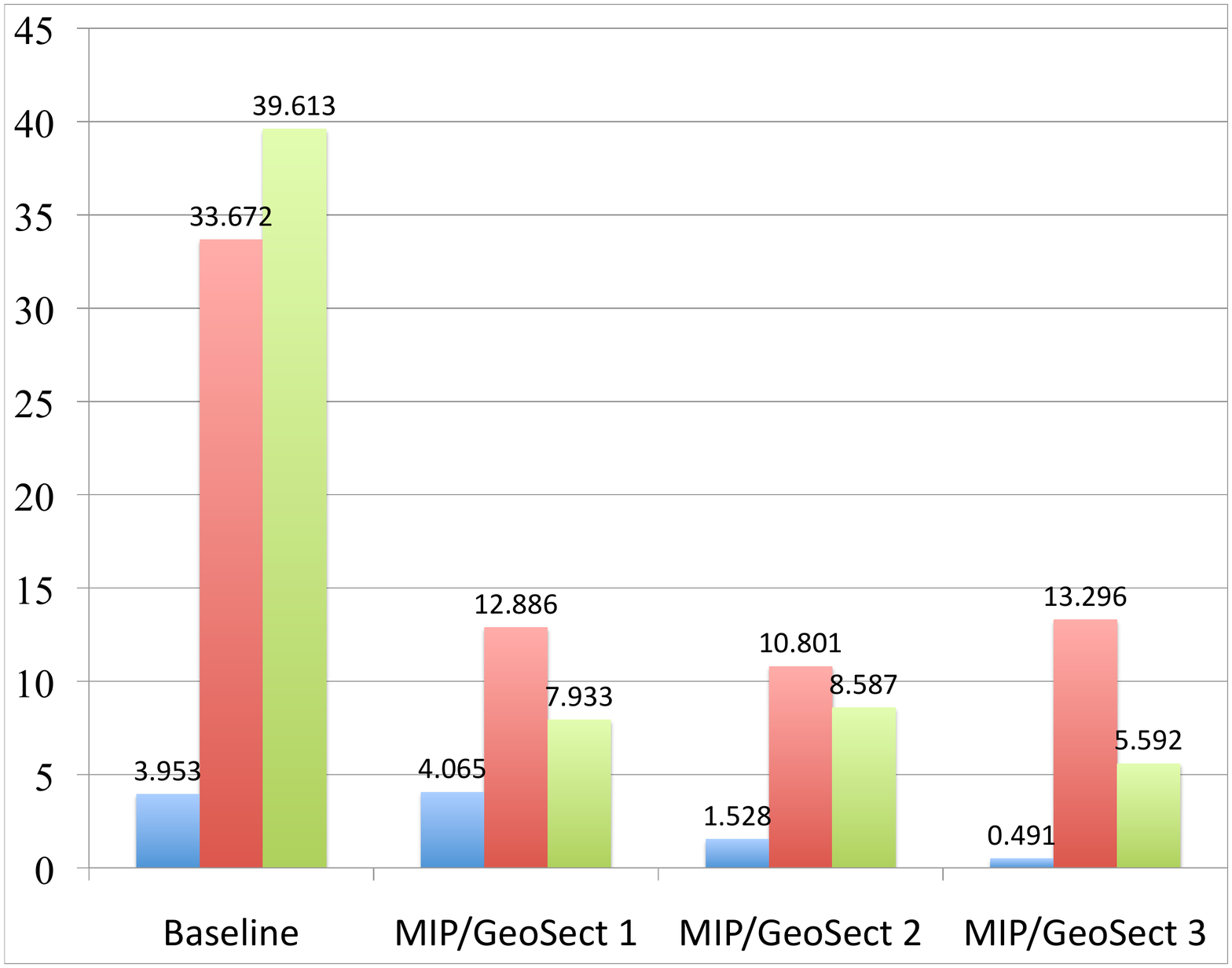}%
    \label{fig:aces_stdev_wl}}}
\caption{\footnotesize Comparison of flight delays induced by baseline sectorizations and three sets of MIP/GeoSect sectorizations.}
\label{fig:aces}
\end{figure}

Our colleagues at Metron Aviation used the Mixed Integer Programming (MIP) method \cite{YD2004} to generate the seed sectorizations that were processed with \GeoSectLocal. Our objective was to minimize the estimated delay and to compare how well it corresponds to the delay computed by the NASA's ACES Flight simulator. We have generated three different sets of sectorizations with the same number of sectors or less than in the baseline sectorizations. The comparison of the delays in the resulting MIP/\GeoSectLocal sectorizations is presented in Figure~\ref{fig:aces}. MIP/\GeoSectLocal approach was able to almost eliminate the average delay in the first stage, and reduce it in $2.5$-$7.8$ times for the second and third stages.

\section{Conclusion}
We have studied the \asp problem, reviewed its complexity and presented a heuristic \algo that uses a multi-criteria optimization approach to improve an input sectorization. 

We have implemented \algo algorithms in \GeoSectLocal, a highly configurable tool that produces high quality sectorizations. The result  depends on the choice of initial sectorizations (more specifically on their topology, and to a less degree on their exact geometry). In all the experiments \GeoSectLocal significantly reduced the average aircraft count and delay, and the sectors had ``nice'' geometric and flow conforming properties. One of the important advantages of the \algo is that it can optimize sectorizations with respect to any constraint that can be described with a simple parameter that can be evaluated numerically. Thus, \algo is not limited to the set of constraints described in this paper.

Current and future work include extending \algo to produce dynamic sectorizations that continuously adapt to the changing traffic, and to produce robust sectorizations under uncertain traffic patterns.

\section*{Acknowledgment}

This work is supported by NASA Ames research Center under contracts number NNA07BB33C and number NNA11AB99C.

We thank our colleagues Girishkumar Sabhnani and Arash Yousefi from Metron Aviation for useful discussions and domain expertise on air traffic management. We also thank Metron Aviation for providing us with the generated sectorizations, and historical and generated track data.

\bibliographystyle{IEEEtran}
\bibliography{IEEEabrv,sea}

\end{document}